# Enhanced superconductivity with possible re-appearance of charge density wave states in polycrystalline $Cu_{1-x}Ag_xIr_2Te_4$ alloys

Mebrouka Boubeche[a,#], Lingyong Zeng[a,#], Xunwu Hu[b], Shu Guo[c,d], Yiyi He[a], Peifeng Yu[a], Yanhao Huang[a], Chao Zhang[a], Shaojuan Luo[e], Dao-Xin Yao[b], Huixia Luo[a*]

[a] School of Materials Science and Engineering, State Key Laboratory of Optoelectronic Materials and Technologies, Key Lab of Polymer Composite & Functional Materials, Guangzhou Key Laboratory of Flexible Electronic Materials and Wearable Devices, Sun Yat-Sen University, No. 135, Xingang Xi Road, Guangzhou, 510275, P. R. China

[b] School of Physics, State Key Laboratory of Optoelectronic Materials and Technologies, Sun Yat-Sen University, No. 135, Xingang Xi Road, Guangzhou, 510275, P. R. China

[c] Shenzhen Institute for Quantum Science and Engineering, Southern University of Science and Technology, Shenzhen 518055, China.

[d] International Quantum Academy, Shenzhen 518048, China.

[e] School of Chemical Engineering and Light Industry, Guangdong University of Technology, Guangzhou, 510006, P. R. China

[#] These authors contributed equally to this work.
[*] Author to whom any correspondence should be addressed.
E-mail: luohx7@mail.sysu.edu.cn

**Abstract**

In this study, we determined the effect of doping with the noble metal Ag on $CuIr_2Te_4$ superconductors. Based on the resistivity, magnetization, and heat capacity, we explored the changes in the superconductivity (SC) and charge density wave (CDW) for $Cu_{1-x}Ag_xIr_2Te_4$ ($0 \leq x \leq 1$) as a function of isoelectric substitution (Cu/Ag). We assessed a complete set of competing states from suppressed CDW in the low doping region to superconductor in the middle doping region and re-entrant CDW in the high doping region, thereby obtaining an electronic phase diagram, where the superconducting dome was near bipartite CDW regions with a maximum superconducting temperature ($T_c$) of about 2.93 K at an Ag doping level of 12%. The lower $H_{c1}$ and upper $H_{c2}$ critical magnetic fields were determined for some representative samples in the $Cu_{1-x}Ag_xIr_2Te_4$ series based on magnetization and resistivity measurements, respectively. We showed that $H_{c1}$ decreased whereas $H_{c2}$ increased as the doping content increased. The specific heat anomalies at the superconducting transitions $\Delta C_{el}/\gamma T_c$ for representative samples comprising $Cu_{0.92}Ag_{0.08}Ir_2Te_4$, $Cu_{0.88}Ag_{0.12}Ir_2Te_4$, and $Cu_{0.85}Ag_{0.15}Ir_2Te_4$ were approximately 1.40, 1.44, and 1.42, respectively, which are all near the Bardeen-Cooper-Schrieffer (BCS) value of 1.43 and they indicate bulk SC in these compounds.

## 1. Introduction

Most of the unconventional high-temperature superconductors (HTSs) are characterized by doping dependence of the superconducting dome close to a competing ordered phase, such as a spin density wave (SDW), charge density wave (CDW), or antiferromagnetic order [1-6]. A widely accepted explanation for this behavior is critical phase fluctuations in the intertwined electronic order. Despite many clear experimental results, the nature of these superconducting domes is still under investigation. A superconducting dome is not a general feature of conventional low-temperature superconductors, but it has been observed in several systems, including Fe with superconducting temperatures ($T_c$) below 2 K at pressures between 15 and 30 GPa [7], and a gated $LaAlO_3/SrTiO_3$ interface [8] doped with $SrTiO_3$ [9,10].

The CDW state is a quantum mechanical phenomenon, which is usually accompanied by a periodic distortion of the lattice first predicted by Peierls [11], and it occurs in a wide range of materials, including the aforementioned HTSs and low dimensional transition metal dichalcogenides (TMDs). The interplay between CDW and the other electronic states of these materials is technologically important, and it has attracted much attention from the nanoelectronics community. CDW materials can be tuned with various parameters (e.g., chemical doping, physical pressure, and gating) to explore the links between various electronic orders [12,13]. In particular, a superconducting dome is commonly a function of a tuning parameter in the proximity of CDW in many TMDs [14-22]. Similar to high $T_c$ cuprates or iron-based superconductors, the same feature with a superconducting dome around the collapse of CDW is also observed in chemical-doped TMDs, which is considered to provide a basis for understanding the mechanism in unconventional HTSs. Thus, studying the interaction between the CDW and superconductivity (SC) in TMDs remains a key focus.

SC usually appears or it is enhanced when the CDW is drastically suppressed by disorder, intercalation, or pressure in the family of low-dimensional layered TMDs. In particular Li et al. [23] gradually substituted sulfur (S) for selenium (Se) in $2H$-$TaSe_2$ to obtain $2H$-$TaS_2$ and robust superconducting order was observed in the single crystal $TaSe_{2-x}S_x$ ($0 < x < 2$) alloy. The $T_c$ values of these $TaSe_{2-x}S_x$ series are much higher than those of the two undoped compounds $TaSe_2$ and $TaS_2$, and the conductivity is higher near the middle of the alloy series compared with 2H-TaSe$_2$ and 2H-TaS$_2$, thereby indicating that SC competes with CDW in this system. In addition, further calculations showed that the disorder facilitates the SC state at the expense of CDW order, as also observed in the experiments mentioned above, according to the real-space self-consistent Bogoliubov–de Gennes calculations and momentum-space calculations involving density functional theory and dynamical mean field theory [24]. Another study demonstrated the coexistence of CDW and SC in 2H-TaS$_2$ at low temperatures by applying hydrostatic pressures. A superconducting dome is observed with a maximum of $T_c$ = 9.1 K and the CDW is suppressed under compression. These calculations indicate that an electronic topological transition occurs before the suppression of phonon instability, thereby suggesting that the electronic topological transition alone does not directly initiate the structural change in 2H-TaS$_2$ [12]. A recent study of the effects of pressure on the CDW and SC in the NbSe$_2$ and NbS$_2$ systems showed that the rapid destruction of the CDW under pressure in NbSe$_2$ is due to quantum fluctuations when the lattice is renormalized by the anharmonic part of the lattice potential. However, based on the analogous superconducting gaps for both NbSe$_2$ and NbS$_2$, the CDW does not affect the superconducting gap structure [13].

Recently, the coexistence of SC and CDW was observed in the quasi-two-dimensional (2D) CuIr$_2$Te$_4$ [25], which has a NiAs defected structure with the trigonal symmetry space group *P*3-*m*1 [25,26], a superconducting transition temperature of $T_c$ = 2.5 K, and a CDW-like transition $T_{CDW}$ occurs at 186 K from cooling and 250 K from warming [25]. Chemical doping is an effective method for exploring new superconductors or tuning the physical properties of existing superconductors. We previously demonstrated that electron dopants (e.g., 3*d* Zn) [27] or hole dopants (e.g., 3*d* Ti; 4*d* Ru) [28,29] can suppress the CDW order, but the superconducting phase diagram was quite different. For example, Cu$_{1-x}$Zn$_x$Ir$_2$Te$_4$ variants doped with electrons exhibited robust SC in the whole doping range of $0 \leq x \leq 0.9$, whereas the CuIr$_{2-x}$(Ti/Ru)$_x$Te$_4$ series doped with holes produced a dome-shaped superconducting phase diagram, where the maximum $T_c$ values were around 2.79 K and 2.84 K respectively [27-29]. However, this trend does not apply to 3*p*-5*p* dopants [30,31]. For example, iodine doping produced dome-shaped SC associated with suppression at a very low doping content but the unexpected re-appearance of CDW occurred under high doping [29]. Moreover, silver (Ag) has been used widely as a dopant or additive to improve the superconducting properties of

HTSs because Ag can improve the inter-grain connections as well as enhancing the critical current and irreversibility field [32,33].

In this study, we chemically substituted Ag ($4d$) for Cu ($3d$), which is below Cu ($3d$) in the same column of the periodic table. We investigated the effects of substituting the non-magnetic noble metal Ag ($4d$) for Cu ($3d$) on the structural and physical properties of the CuIr$_2$Te$_4$ series. We identified several positive effects, including enhancement of the critical parameters such as $T_c$ and H$_{c2}$, and a comprehensive electronic phase diagram was constructed to illustrate the overall behavior of $T_c$ and CDW for our samples.

## 2. Experimental details

Polycrystalline samples of Cu$_{1-x}$Ag$_x$Ir$_2$Te$_4$ ($0 \leq x \leq 1$) were prepared using the conventional solid state reaction technique. Stoichiometric mixtures of Cu powder (99%, ~325 mesh, Alfa Aesar), Ag powder (99.9%, ~325 mesh, Alfa Aesar), Ir powder (99.9%, Macklin), and Te lump (99.999%, Alfa Aesar) were sealed in quartz tubes and heated at 850°C for 5 days. The as-prepared powders were ground, pelletized, and annealed in evacuated quartz tubes at 800°C for 10 days. Cu$_{1-x}$Ag$_x$Ir$_2$Te$_4$ ($0 \leq x \leq 1$) samples are stable in the air. Powder X-ray diffraction (XRD) was conducted using a standard diffractometer (MiniFlex, Rigaku apparatus) with Cu K$\alpha$ ($\lambda = 1.5406$ Å) radiation to probe the phase purity. The elemental ratios and distributions in samples were analyzed using a scanning electron microscope (EM-30AX PLUS, Kurashiki Kako Co. Ltd, Japan) equipped with an energy dispersive X-ray spectroscopy detector. Electrical transport, dc susceptibility, and specific heat measurements were performed at low temperatures (down to 1.8 K) using a physical properties measurement system (PPMS, Quantum Design). The temperature-dependent electrical resistivity was tested under different magnetic fields $\rho$(H,T) with rectangular samples using the standard four-probe method. The samples were also characterized based on dc-susceptibility measurements obtained with finely ground powders using the AC measurement system (ACMS) model. $T_c$ was estimated conservatively as the intersection of the extrapolated abrupt slope for the susceptibility in the superconducting transition region and the normal state, as well as from the resistivity as the midpoint of the resistivity $\rho$(T) transitions and from specific heat data $T_c$ obtained using the equal area construction method.

The field and temperature dependences of the magnetic susceptibility χ(H,T) were employed to identify the lower critical fields $H_{c1}$. The upper critical fields $H_{c2}$ were determined using resistivity data collected under increased magnetic fields near the superconducting transition.

First-principles calculations conducted using the projector augmented wave method [34] implemented in the VASP package [35], as described in detail in a previous study [25].

## 3. Results and discussion

The XRD patterns obtained at room temperature for the polycrystalline $Cu_{1-x}Ag_xIr_2Te_4$ (0 ≤ x ≤ 1) samples are shown in **Fig. 1(a–b)**. As shown in **Fig. 1(a)**, the (*h k l*) reflection peaks indexed well to a trigonal structure with the *P*-3*m*1 space group, except for the peak denoted by * in **Fig. 1(b)** at 2θ = 40.86 °, which was due to the small amount of unreacted Ir. The foreign Ir phase weakened as the amount of Ag doping increased, thereby indicating the enhanced homogeneity and purity of the Ag-doped samples. Thus, the Ag substitution process did not change the crystal structure of the parent material in the substitution range from 0 ≤ *x* ≤ 1. The lattice parameters were calculated for the $Cu_{1-x}Ag_xIr_2Te_4$ samples by applying the pseudo-Voigt function for peak profile fitting and the structural model using FullProf software [36]. The lattice parameter values obtained for the representative sample $Cu_{0.88}Ag_{0.12}Ir_2Te_4$ were *a* = *b* = 3.94155(3) Å, *c* = 5.4122(2) Å with the fitting parameters $R_{wp}$ = 3.3%, $R_{exp}$ = 2.04%, and $R_p$ = 2.85%, and $\chi^2$ = 3.41. **Table 1** shows the detailed Rietveld refinement results for the representative samples $Cu_{0.88}Ag_{0.12}Ir_2Te_4$, $Cu_{0.5}Ag_{0.5}Ir_2Te_4$, and $AgIr_2Te_4$, and details for the other compositions are given in **Table S1** in the supplemental information. We detected a slight shift in the diffraction peaks toward lower angles (see the right panel in **Fig. 1(b)**) as the amount of Ag increased in the samples. Ag is located below Cu in the periodic table, so the Ag atom has a larger radius compared with the Cu atom. Thus, both the lattice constants *a* = *b* and *c* increased with the Ag concentration, as shown in **Fig**. **1(c)**.

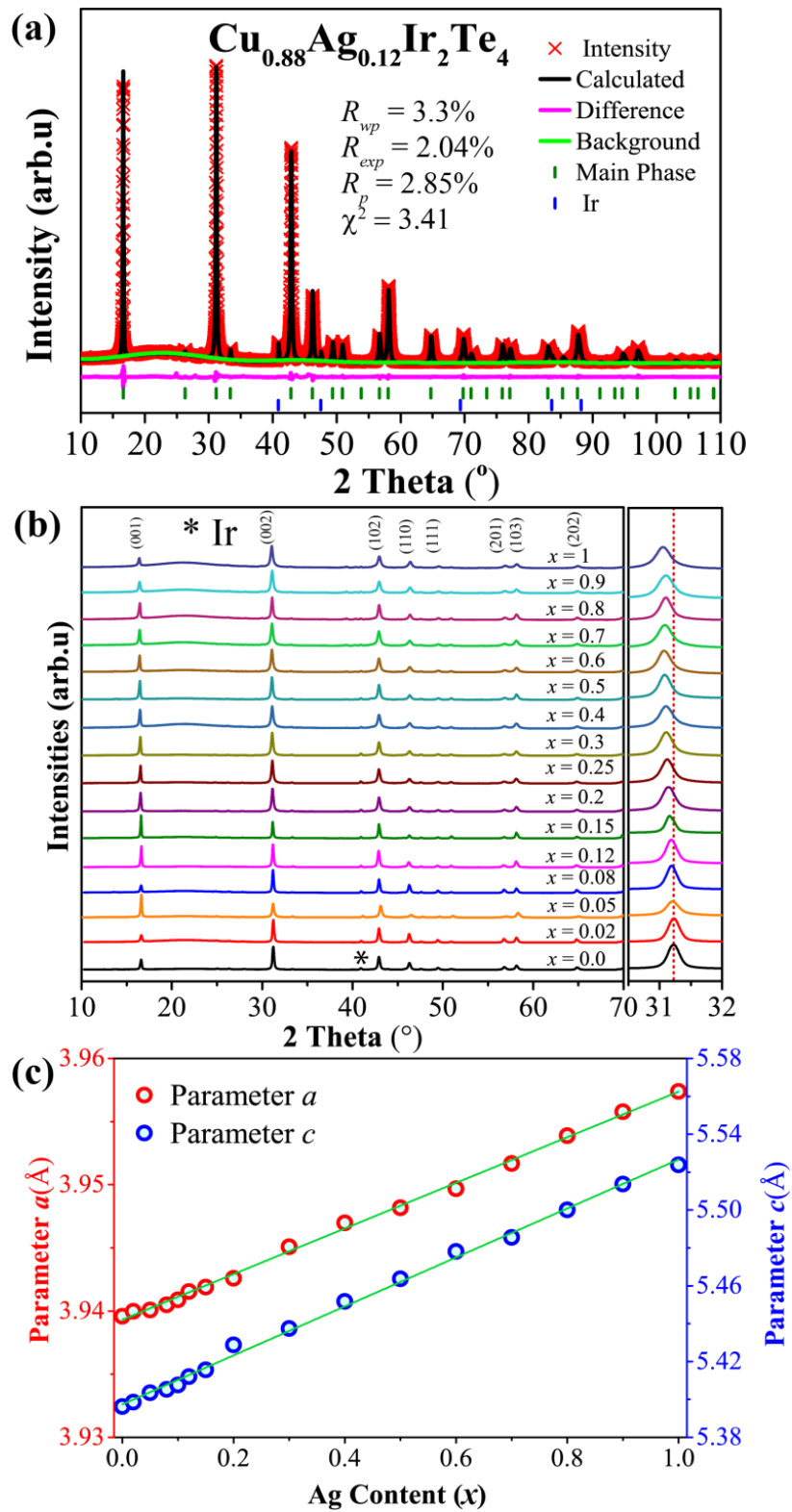

**Fig. 1** Structural analysis of polycrystalline $Cu_{1-x}Ag_xIr_2Te_4$ powders prepared at 300 K. **(a)** Rietveld refinement results for the representative sample $Cu_{0.88}Ag_{0.12}Ir_2Te_4$. **(b)** X-ray diffraction (XRD) pattern obtained for the $Cu_{1-x}Ag_xIr_2Te_4$ polycrystals, where the impurity Ir line is denoted by *. **(c)** The unit cell size for the $Cu_{1-x}Ag_xIr_2Te_4$ system depended on the Ag doping amount.

**Table 1.** Rietveld refinement structural parameters for $Cu_{0.88}Ag_{0.12}Ir_2Te_4$, $Cu_{0.5}Ag_{0.5}Ir_2Te_4$, and $AgIr_2Te_4$ based on the *P-3m1* space group (No. 164).

| $Cu_{0.88}Ag_{0.12}Ir_2Te_4$ | | | | \multicolumn{3}{l}{$R_{wp}$ = 3.3%, $R_p$ = 2.85%, $R_{exp}$ = 2.04%, $\chi^2$ = 3.41} |
|---|---|---|---|---|---|---|
| **Label** | **x** | **y** | **z** | **Site** | **Occupancy** | **Multiplicity** |
| Cu | 0 | 0 | 0.5 | 2b | 0.44 | 1 |
| Ag | 0 | 0 | 0 | 2b | 0.06 | 1 |
| Ir | 0 | 0 | 0 | 1a | 1 | 1 |
| Te | 0.33333 | 0.66667 | 0.74508(1) | 2b | 1 | 2 |
| $Cu_{0.5}Ag_{0.5}Ir_2Te_4$ | | | | \multicolumn{3}{l}{$R_{wp}$ = 3.45%, $R_p$ = 2.48%, $R_{exp}$ = 2.1%, $\chi^2$ = 3.5} |
| **Label** | **x** | **y** | **z** | **Site** | **Occupancy** | **Multiplicity** |
| Cu | 0 | 0 | 0.5 | 2b | 0.25 | 1 |
| Ag | 0 | 0 | 0.5 | 2b | 0.25 | 1 |
| Ir | 0 | 0 | 0 | 1a | 1 | 1 |
| Te | 0.33330 | 0.66667 | 0.74671(6) | 2b | 1 | 2 |
| $AgIr_2Te_4$ | | | | \multicolumn{3}{l}{$R_{wp}$ = 3.88%, $R_p$ = 3.42%, $R_{exp}$ = 2.11%, $\chi^2$ = 3.66} |
| **Label** | **x** | **y** | **z** | **SITE** | **Occupancy** | **Multiplicity** |
| Ag | 0 | 0 | 0.5 | 2b | 0.5 | 1 |
| Ir | 0 | 0 | 0 | 1a | 1 | 1 |
| Te | 0.33330 | 0.66667 | 0.76405(3) | 2b | 1 | 2 |

We conducted scanning electron microscopy-energy dispersive X-ray spectroscopy analysis to determine the distributions and ratios of the elements in the $Cu_{1-x}Ag_xIr_2Te_4$ system. The crystalline structures of the samples are shown in **Fig. S1**, which indicates that the grain size decreased as the Ag content increased. In addition, the elements were uniformly distributed in the $Cu_{1-x}Ag_xIr_2Te_4$ polycrystalline samples (see **Fig. S2**). The experimental percentages of Cu:Ag:Ir:Te in these samples were close to the target percentages (see **Fig. S3** and **Table S2**).

**Figure 2(a–b)** shows the temperature-dependent resistivities ($\rho/\rho_{300K}$) normalized relative to their corresponding 300 K values for the samples investigated in the temperature range from 1.8–300 K under zero applied field. The resistivities of the samples with $x \leq 0.15$ decreased as the temperature decreased from 300 to 3 K, thereby indicating the metallic behavior of the samples investigated. We found that the resistivities of these samples could be fitted by $\rho \sim T^2$ in the low-temperature range, and thus they exhibited Fermi-liquid behavior (see the blue line in the inset in **Fig. 2(a)**). In addition, the resistivities of the samples with doping amounts of $0 \leq x \leq 0.3$ suddenly decreased to zero at low temperatures due to the appearance of SC as the temperature decreased. The hump-like anomaly observed for the parent compound $CuIr_2Te_4$ around 180 K was rapidly suppressed by adding a small amount of Ag ($x = 0.02$). However,

the resistivity anomaly linked to the CDW order re-appeared again in the high doping area when $x \geq 0.2$ and it shifted to higher temperatures as the Ag content increased (see **Fig. 2(b)** and the inset, which shows that the minimum of $d\rho/dT$ determines $T_{CDWS}$). Similar behavior was also observed previously in an iodine and selenium doped system [30,31], which was caused by the disorder due to doping.

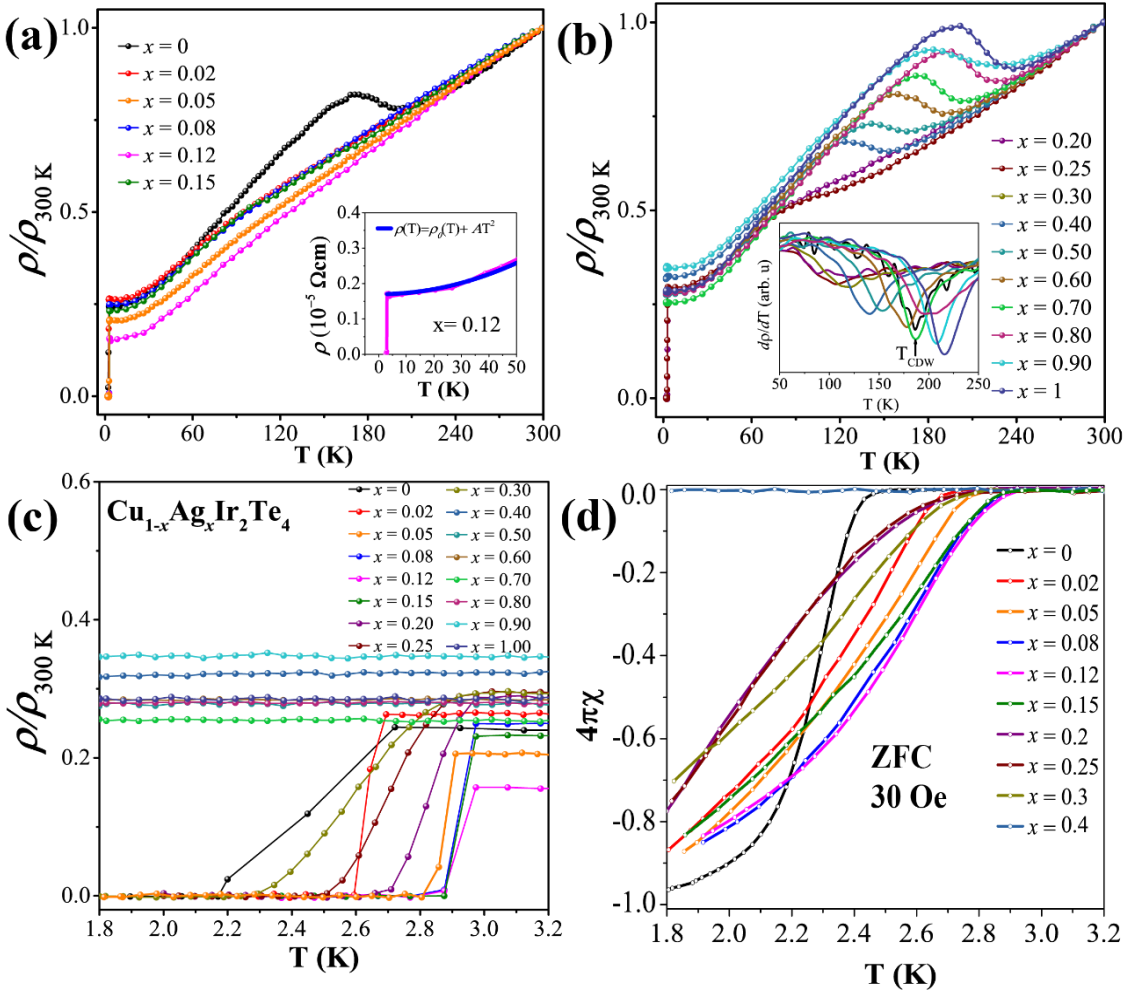

**Fig. 2** Electrical and magnetic characteristics of the $Cu_{1-x}Ag_xIr_2Te_4$ system. **(a)** Normalized resistivity $\rho/\rho_{300\,K}$ (T) curves below 300 K down to 1.8 K for $Cu_{1-x}Ag_xIr_2Te_4$ ($0 \leq x \leq 0.15$), where the inset shows the temperature dependence of $\rho(T)$ up to 300 K for the $Cu_{0.88}Ag_{0.12}Ir_2Te_4$ sample, which was fitted by $\rho \sim T^2$ (solid blue line). **(b)** Normalized resistivity $\rho/\rho_{300\,K}$ (T) curves below 300 K down to 1.8 K for $Cu_{1-x}Ag_xIr_2Te_4$ ($0.2 \leq x \leq 1$). The inset shows the first derivative of the resistivity $d\rho/dT$ (T) curve at high temperatures measured from cooling, which was used to evaluate the CDW-like transition temperature ($T_{CDW}$). **(c)** Normalized resistivity $\rho/\rho_{300\,K}$ (T) curves below 3.3 K down to 1.8 K. **(c)** Normalized dc magnetic susceptibility $\chi(T)$ in the zero field cooled (ZFC) regime at 30 Oe for $Cu_{1-x}Ag_xIr_2Te_4$ ($0 \leq x \leq 0.4$).

**Figure 2(c)** illustrates the temperature dependence of the normalized resistivity below 3.3 K. The samples in the range from 0 to 0.3 exhibited a sharp superconducting transition. The $T_c$ values were determined from the mid-point of the normalized resistivity during the decrease in resistance for the SC. Interestingly, $T_c$ was enhanced from 2.5 K for the undoped sample to 2.93 K for $x = 0.12$. Subsequently, $T_c$ decreased gradually with further Ag doping. For $x \geq 0.4$, no signs of SC were observed above 1.8 K. **Table 2** and **Fig. S5** show that the increase in $T_c$ was followed by an increase in the residual resistivity ratio (RRR = $R_{300K}/R_{5K}$) from 3.56 for the host sample to 6.68 for the optimal composition ($x = 0.12$) with the highest $T_c$, but the RRR values suddenly decreased for $x > 0.15$ to 2.94 for $x = 1$ (see **Table 2** and **Fig. S5**). It should be noted that Ag-doped samples in the range from $0 < x \leq 0.2$ exhibited sharp superconducting transitions, thereby suggesting that these compounds were highly homogeneous. The decrease in RRR suggests that Ag doping significantly induced disorder and that Ag ions were effective scattering centers [37-39], which could explain the re-emergence of the CDW. **Figure 2(d)** shows the normalized zero-field cooling regime dc magnetic susceptibility $\chi(T)$ at H = 30 Oe in the vicinity of $T_c$ for $Cu_{1-x}Ag_xIr_2Te_4$ ($0 \leq x \leq 0.4$). The samples with $0 \leq x \leq 0.3$ produced diamagnetic signals below the superconducting transition temperature, and the superconducting shielding fraction was around 72–96%. The $T_c$ values extracted from the magnetic susceptibility measurements were consistent with those obtained from the electrical measurements, as shown in **Fig. 2(c)**.

**Table 2.** Dependences of the residual resistance ratio (RRR = $R_{300K}/R_{5K}$) on the Ag content ($x$) superconducting transition temperature ($T_c$), and CDW transition temperature ($T_{CDW}$).

| Ag content ($x$) | RRR = $R_{300K}/R_{5K}$ | $T_c$ (K) | $T_{CDW}$ (K) |
|---|---|---|---|
| 0 | 3.56 | 2.5 | 187 |
| 0.02 | 3.85 | 2.63 | |
| 0.05 | 5 | 2.87 | |
| 0.08 | 4.4 | 2.92 | |
| 0.12 | 6.68 | 2.93 | |
| 0.15 | 4.35 | 2.91 | |
| 0.2 | 3.57 | 2.83 | 87 |
| 0.25 | 3.44 | 2.7 | 105 |
| 0.3 | 3.42 | 2.59 | 120 |
| 0.4 | 3.40 | 1.8 | 140 |
| 0.5 | 3.38 | | 153 |
| 0.6 | 3.42 | | 178 |
| 0.7 | 3.8 | | 186 |
| 0.8 | 3.42 | | 207 |

| | | |
|---|---|---|
| 0.9 | 2.92 | 207 |
| 1 | 2.94 | 216 |

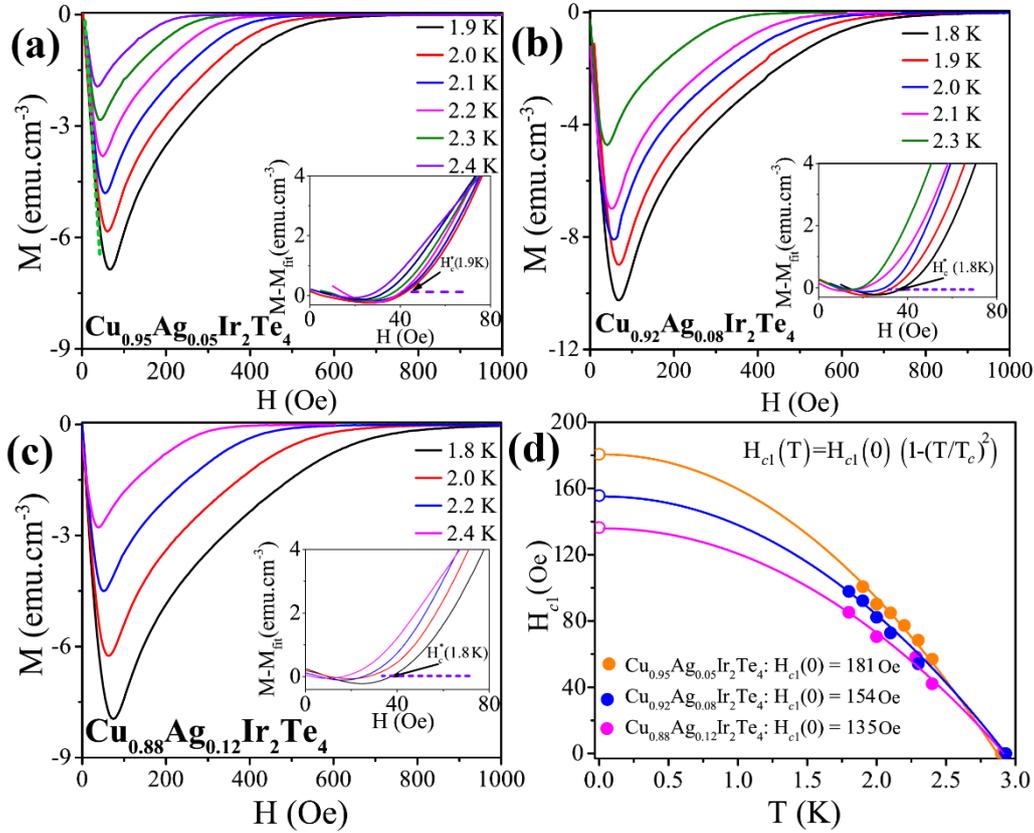

**Fig. 3** Measurements of lower critical field $H_{c1}$. **(a, b, c)** Field-dependent magnetization isotherms for $Cu_{0.95}Ag_{0.05}Ir_2Te_4$, $Cu_{0.92}Ag_{0.08}Ir_2Te_4$, and $Cu_{0.88}Ag_{0.12}Ir_2Te_4$, respectively. The insets show the corresponding $M-M_{fit}(H)$ at different temperatures. **(d)** Lower critical field fittings for $Cu_{0.95}Ag_{0.05}Ir_2Te_4$ (orange), $Cu_{0.92}Ag_{0.08}Ir_2Te_4$ (blue), and $Cu_{0.88}Ag_{0.12}Ir_2Te_4$ (pink).

The lower and upper critical fields ($H_{c1}$ and $H_{c2}$) were determined to provide a comprehensive overview of the mixed state characteristics of the type-II superconductors. $H_{c1}$ is defined as the field where the magnetization curve begins to deviate from the linear track. $H_{c1}$ is linked to the free energy of the flux lines and it provides information concerning the crucial mixed state properties, such as the Ginzburg–Landau parameter ($\kappa$) and the penetration depth ($\lambda$). **Figure 3(a–c)** shows the field-dependent magnetization isotherms M(H) obtained at several temperatures for representative samples $Cu_{0.95}Ag_{0.05}Ir_2Te_4$, $Cu_{0.92}Ag_{0.08}Ir_2Te_4$, and $Cu_{0.88}Ag_{0.12}Ir_2Te_4$. The shapes of the M(H) curves suggest that $Cu_{0.95}Ag_{0.05}Ir_2Te_4$, $Cu_{0.92}Ag_{0.08}Ir_2Te_4$, and $Cu_{0.88}Ag_{0.12}Ir_2Te_4$ are type-II superconductors, as indicated by the linear shielding ("Meissner line") at low fields (see the dashed green line in the main panels). Above

~ 100 Oe, the shielding reduced as the magnetic flux started to penetrate the bulk and the system entered the vortex state. To determine $H_{c1}(0)$, we used the most popular method for different superconducting systems [40], as shown in the insets in **Fig. 3(a–c)**. The linear green dashed straight line in **Fig. 3(a)** denotes the Meissner shielding effects at low fields fitted using the Meissner line formula: $M_{fit} = A + BH$, where $A$ and $B$ are the intercept and the slope of the linear fitting for M(H) data at a low magnetic field, respectively. We subtracted the Meissner line obtained based on the low field magnetization slope from the magnetization (M) for each isotherm M-$M_{fit}$ (H). The value of $H_{c1}$* was estimated from 1% M when it diverged from the fitted data ($M_{fit}$), as shown by the horizontal dashed purple lines in the insets in **Fig. 3(a–c)**. To accurately estimate $H_{c1}$, we considered the demagnetization effects. The demagnetization factor ($N$) can be predicted from the formula: $N = 1/(4\pi\chi_V + 1)$, where $\chi_V = dM/dH$ is the value of the fitted slope, as shown in **Fig. 3(a–c)**. The $N$ values were obtained as around 0.58, 0.56, and 0.53 for $Cu_{0.95}Ag_{0.05}Ir_2Te_4$, $Cu_{0.92}Ag_{0.08}Ir_2Te_4$, and $Cu_{0.88}Ag_{0.12}Ir_2Te_4$, respectively. The $H_{c1}$ values obtained versus temperature are plotted in **Fig. 3(d)**. These curves were fitted using the overall equation for $H_{c1}$: $H_{c1}(T) = H_{c1}(0)(1-(T/T_c)^2)$, i.e., $H_{c1}(0)$ was estimated from the extrapolation of the $H_{c1}(T)$ data down to 0 K. The color online open symbols in **Fig. 3(d)** show that $H_{c1}(0)$ = 181 Oe, 154 Oe, and 135 Oe for $Cu_{0.95}Ag_{0.05}Ir_2Te_4$, $Cu_{0.92}Ag_{0.08}Ir_2Te_4$, and $Cu_{0.88}Ag_{0.12}Ir_2Te_4$, respectively, which are all lower than that for the pristine $CuIr_2Te_4$ (280 Oe) [25], as shown in **Table 3**.

**Figure 4(a–c)** show the temperature-dependent resistivities measured by applying different magnetic fields to $Cu_{0.88}Ag_{0.12}Ir_2Te_4$, and $Cu_{0.8}Ag_{0.2}Ir_2Te_4$. The magnetic field had no substantial effects on transition broadening. A sharp transition appeared even at higher fields. $T_c(H)$ was taken from the 10%, 50%, and 90% criteria (see **Fig. S5** in the supplemental information), which allowed us to fit the $H_{c2}(T)$ phase diagrams for the selected compounds, as shown in **Fig. 4(b–d)**. The best fits of the experimental data were obtained using the dirty limit equation [31]: $h^*_{fit} = 1 - t - C_1(1-t)^2 - C_2(1-t)^4$ where $t$ is the reduced temperature, $C_1$ = 0.153, and $C_2$ = 0.152, which was shown by Baumgartner et al. to provide an excellent estimation of the dirty limit temperature-dependent upper critical field according to Werthamer, Helfand, and Hohenberg theory [41,42]. Using this equation, the upper critical field at any temperature below $T_c$ can be obtained using the formula: $H_{c2}(T) = \dfrac{H_c(0)}{0.693} h^*_{fit}(T/T_c)$ [41,42]. The $H_{c2}(0)$ values obtained for the $Cu_{0.88}Ag_{0.12}Ir_2Te_4$ and $Cu_{0.8}Ag_{0.2}Ir_2Te_4$ samples based on the 50% criteria were 2100 and 1400 Oe, respectively. These two samples exhibited a slight increase in the upper critical field compared with the pristine $CuIr_2Te_4$ (see **Table 3**). The $H_{c2}(0)$ values are lower than the Pauli limiting field for weak-coupling Bardeen-Cooper-Schrieffer (BCS) superconductors estimated from $H^P = 1.86*T_c$ T/K [44], which are around 5.4 T and 5.2 T for $x = 0.12$ and $x = 0.2$, respectively. The $H^P$ values obtained for the doped samples were greater than that for the parent $CuIr_2Te_4$ ($H^P = 4.65$ T) due to their higher $T_c$ values. The Ginzburg–Landau coherence length at 0 K ($\xi_{GL}(0)$) was calculated using the following equation: $H_{c2} = \phi_0/(2\pi\xi_{GL}^2)$ [45], where $\phi_0 = 2.07 \times 10^{-3}$ T $\mu m^2$ is the flux quantum. The $\xi_{GL}(0)$ values were determined for $Cu_{0.88}Ag_{0.12}Ir_2Te_4$ and $Cu_{0.8}Ag_{0.2}Ir_2Te_4$ as 40 and 48 nm, respectively.

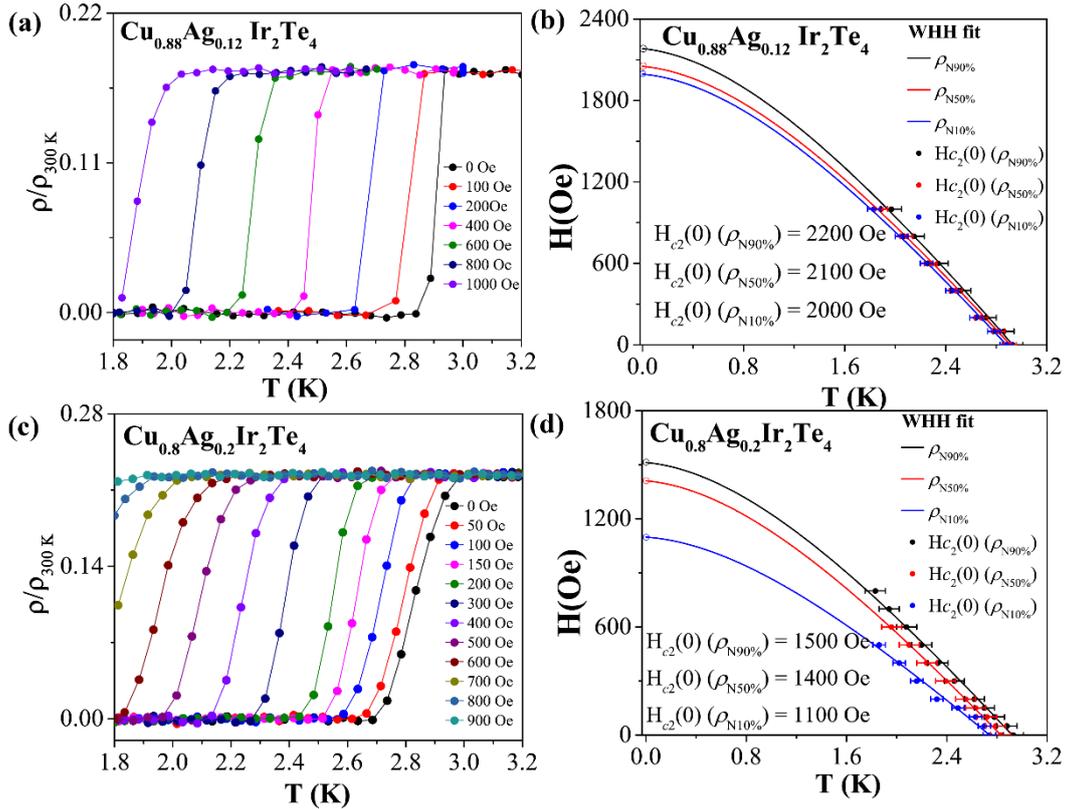

**Fig. 4** Measurements of the upper critical field $H_{c2}$. **(a, b)** Field-dependent magneto-resistivity and $H_{c2}$ phase diagram for $Cu_{0.88}Ag_{0.12}Ir_2Te_4$. **(c, d)** Field-dependent magneto-resistivity and $H_{c2}$ phase diagram for $Cu_{0.8}Ag_{0.2}Ir_2Te_4$.

The lower critical field $H_{c1}$ is linked to the coherence length $\xi$ and magnetic penetration depth $\lambda$ by the relationship: $H_{c1} = (\phi_0/4\pi\lambda^2) (\ln(\kappa) + 0.5)$ [43], where $\kappa = \lambda/\xi$ is the Ginzburg–

Landau parameter. For the optimal compound with $x = 0.12$, we obtained $\lambda = 121$ nm and $\kappa = 3.05$, which correspond to a type II superconductor ($\kappa > 1/\sqrt{2}$) [46].

The temperature-dependent specific heat $C_p(T)$ values at zero magnetic field (0 Oe) for $Cu_{0.92}Ag_{0.08}Ir_2Te_4$ and $Cu_{0.88}Ag_{0.12}Ir_2Te_4$ are shown in **Fig. 5(a)**. The dashed colored lines denote the fit for the expression: $C_p/T(T) = \gamma + \beta T^2$, where $\gamma$ is the electronic specific heat constant and $\beta$ is the phonon contribution term. The fitting characteristics were determined as $\gamma = 12.1 \pm 0.1$ mJ mole$^{-1}$ K$^{-2}$ and $\beta = 1.98 \pm 0.05$ mJ mol$^{-1}$ K$^{-4}$ for $Cu_{0.92}Ag_{0.08}Ir_2Te_4$, $\gamma = 13.9 \pm 0.08$ mJ mole$^{-1}$ K$^{-2}$ and $\beta = 2.12 \pm 0.02$ mJ mol$^{-1}$ K$^{-4}$ for the optimal compound $Cu_{0.88}Ag_{0.12}Ir_2Te_4$, and $\gamma = 14.8 \pm 0.1$ mJ mole$^{-1}$ K$^{-2}$, $\beta = 2.08 \pm 0.05$ mJ mol$^{-1}$ K$^{-4}$ for $Cu_{0.85}Ag_{0.15}Ir_2Te_4$. Thus, the constant $\beta$ was used to obtain the distinct Debye temperature $\Theta_D$ using the formula: $\Theta_D = (12\pi^4 nR/5\beta)^{1/3}$, where $n$ is the number of atoms contained in one unit cell and $R$ is the gas constant. The $\Theta_D$ values were obtained for $Cu_{0.92}Ag_{0.08}Ir_2Te_4$, $Cu_{0.88}Ag_{0.12}Ir_2Te_4$, and $Cu_{0.85}Ag_{0.15}Ir_2Te_4$ as 189 K, 186 K, and 187 K, respectively. The electronic contribution of the specific heat $C_{el}(T)$ measured at H = 0 Oe is shown in the inset in **Fig. 5(a–b)**, which was derived by subtracting the phonon contribution $C_{ph}$. The sharp anomalies comprising the thermodynamic transitions in $Cu_{0.92}Ag_{0.08}Ir_2Te_4$, $Cu_{0.88}Ag_{0.12}Ir_2Te_4$, and $Cu_{0.85}Ag_{0.15}Ir_2Te_4$ from the normal state to the superconducting state were detected at 2.67 K, 2.86 K, and 2.77 K, respectively. Therefore, the compound with $x = 0.12$ had the highest bulk $T_c$ in agreement with the resistivity and magnetic susceptibility data. However, the $T_c$ values were slightly lower than those extracted from $\rho(T)$ and $\chi(T)$ measurements, i.e., 2.91 K ($x = 0.08$) and 2.93 K ($x = 0.12$). This difference in the $T_c$ values is not unusual in CDW materials [47]. The vertical line indicates the equal-area construction (light blue area) at $T_c$ and the linear estimation of the $C_{el}/T$ data immediately above and below $T_c$ (solid orange lines). The specific heat jump was determined from the difference between the $C_{el}$ values at $T_c$ and the normal state, as shown by the vertical solid orange line in the insets in **Fig. 5**. The $\Delta C_{el}/\gamma T_c$ values were determined as 1.40, 1.44, and 1.42 for $Cu_{0.92}Ag_{0.08}Ir_2Te_4$, $Cu_{0.88}Ag_{0.12}Ir_2Te_4$, and $Cu_{0.85}Ag_{0.15}Ir_2Te_4$, respectively. These values are all close to the BCS value of 1.43, thereby suggesting that these compounds are weak coupling superconductors. Using the $\Theta_D$ and $T_c$ values, we obtained the value of the electron–phonon coupling constant ($\lambda_{ep}$) based on the inverted McMillan's expression [46]: $\lambda_{ep} = \frac{1.04 + \mu^* \ln(\Theta_D/1.45T_c)}{((1-0.62\mu^*)\ln(\Theta_D/1.45T_c) - 1.04)}$, where $\mu^* = 0.13$ is the Coulomb pseudopotential, which considers the direct Coulomb repulsion between electrons. The values of $\lambda_{ep}$ were calculated as around 0.62 for both compounds. The Fermi level $N(E_F)$ near the density of states (DOS) can be obtained from the formula: $N(E_F) = 3\gamma/(\pi^2 k_B^2(1+\lambda_{ep}))$, where $k_B$ is Boltzmann's constant. The values of $N(E_F)$ were determined as 3.27 states/eV f.u. for $Cu_{0.92}Ag_{0.08}Ir_2Te_4$, 3.61 states/eV f.u. for $Cu_{0.88}Ag_{0.12}Ir_2Te_4$, and 3.68 states/eV f.u. for $Cu_{0.85}Ag_{0.15}Ir_2Te_4$, which are larger than that for $CuIr_2Te_4$ (see **Table 3**). The increase in the SC in the $Cu_{1-x}Ag_xIr_2Te_4$ compounds can be explained according to the increase in the DOS of the Fermi surface $N(E_F)$, and it may also be related to the enrichment of electron–phonon coupling caused by Ag ion substitution compared with the host $CuIr_2Te_4$ (see **Table 3**). In addition, **Table 3** compares data for the SC and normal states of non-magnetic element-doped $CuIr_2Te_4$ prepared using a solid state reaction method, where it shows that silver and iodine doped systems have the highest $T_c$ values. The $N(E_F)$ and $H_{c2}(0)$ values for the $Cu_{1-x}Ag_xIr_2Te_4$ system are comparable to those for other systems. These compounds are all weak coupling bulk superconductors because their specific heat jump values are close to 1.43.

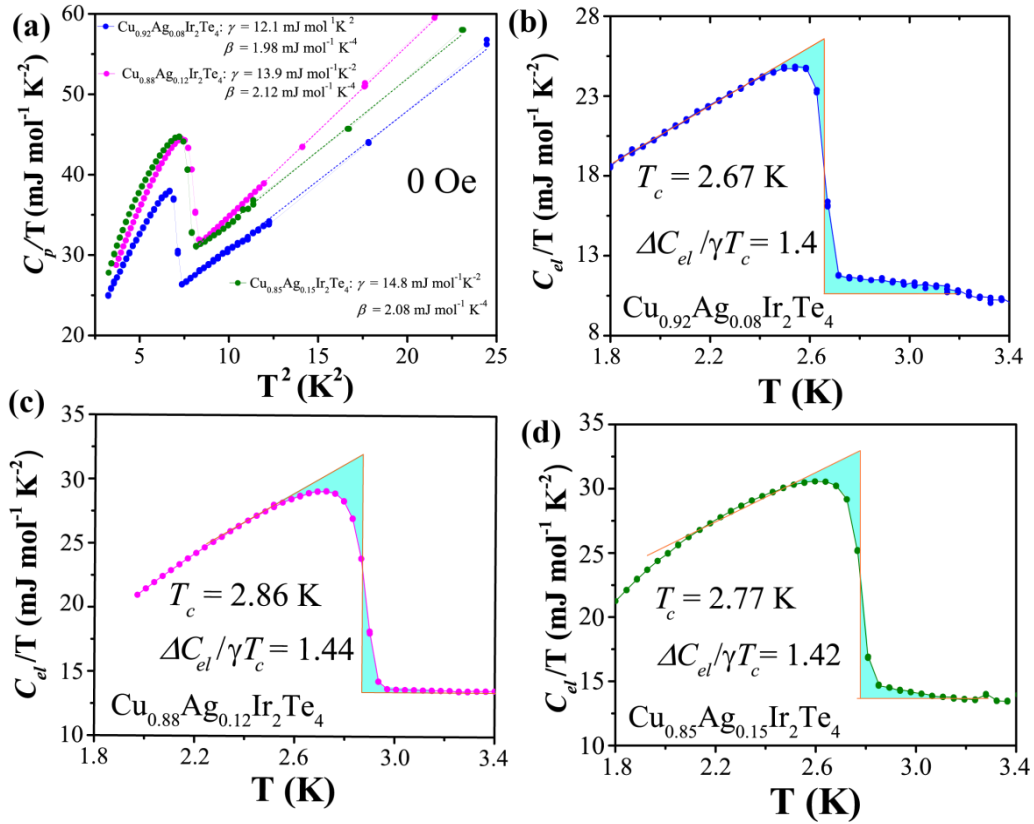

**Fig. 5 (a)** Specific heat versus temperature plot $C_p/T$ ($T^2$) in the temperature range of 2–5 K for $Cu_{0.92}Ag_{0.08}Ir_2Te_4$, $Cu_{0.88}Ag_{0.12}Ir_2Te_4$, and $Cu_{0.85}Ag_{0.15}Ir_2Te_4$, where the dashed lines represent the linear fits $C_p/T(T) = \gamma + \beta T^2$. **(b, c, d)** Electronic specific heat values for $Cu_{0.92}Ag_{0.08}Ir_2Te_4$, $Cu_{0.88}Ag_{0.12}Ir_2Te_4$, and $Cu_{0.85}Ag_{0.15}Ir_2Te_4$, respectively.

Finally, we determined the effects of Ag doping on the SC and CDW in $CuIr_2Te_4$. **Figure 6** summarizes the dependences of $T_c$ and $T_{CDW}$ on $x$. In general, $T_{CDW}$ disappeared with a very low Ag doping concentration, followed by a dome-shaped superconducting phase. In the substitution range from $0 \leq x \leq 0.12$, $T_c$ increased slightly as $x$ increased and reached a maximum of $T_c = 2.93$ K at $x = 0.12$, before decreasing at a medium doping content until full suppression down to 1.8 K at $x = 0.4$. The CDW-like state re-appeared as $T_c$ decreased and $T_{CDW}$ increased, thereby leading to an interesting bipartite phase diagram with a superconducting dome confined between suppressed and re-appeared CDW states. This behavior of the CDW has also been found after substituting other elements such as iodine and selenium in the tellurium site [30,31]. An abnormal hump was not linked to the structural transition confirmed by the temperature-dependent XRD patterns for selenium doped $CuIr_2Te_4$ at 20 K, 100 K, and 300 K [31]. Thus, these results differ from the case of $IrTe_2$, where the resistivity hump was associated with the structural phase [46-48]. Moreover, they differ from the results after substituting Zn in the Cu site or Ru/Ti in Ir site in the parent compound $CuIr_2Te_4$ [27,28,51]. Therefore, we conclude that the doping effect on the CDW-like transition in our system appears to be dopant-dependent. Similar behavior was reported in Tl-intermediate $Nb_3Te_4$ single crystals [50] and it was ascribed to the disorder formed in quasi-one-dimensional Nb chains.

Similarly, the intercalation of a $3d$ transition metal in $M_xTiSe_2$ ($M$ = Mn, Cr, Fe) produces analogous behavior, with suppression initially and then re-emergence of CDW order with the intercalation of higher amounts of $3d$ metals, probably due to the deformation degree of Se-Ti-Se sandwiches [53,54]. The re-appearance of the CDW also occurs in higher doped $1T$-$TaS_{2-x}Se_x$ single crystals [20]. Another example is $2H$-$TaSe_{2-x}S_x$ ($0 \leq x \leq 2$) where CDW occurs simultaneously at two ends, and disorder is an important factor related to the CDW and SC behavior [55]. Based on these similarities, it is reasonable to suggest that the possible re-emergence of the CDW-like transition in $Cu_{1-x}Ag_xIr_2Te_4$ bulks is due to the disorder produced by Ag doping. In addition, RRR universally expresses the disorder in dirty superconductors [37-39]. It is also well known that the RRR is reduced in the dirty-band case [56-59]. However, in our study, the RRR generally increased as the doping content increased in the lower doping region from 0 to 0.12, but decreased as *x* increased at higher doping amounts of $0.12 < x \leq 1$. As shown in **Fig. 6**, $T_c$ increased when the CDW transition was suppressed, whereas $T_c$ decreased when the CDW transition re-emerged and $T_{CDW}$ increased as the doping content increased further. In general, our series exhibited opposing trends toward CDW and SC orders. **Figure S6** shows the band structure for $Cu_{0.9}Ag_{0.1}Ir_2Te_4$. The Fermi level of $Cu_{0.9}Ag_{0.1}Ir_2Te_4$ $E_F$ is 9.187 eV. The vertical axis of the energy band diagram is $E$-$E_F$. **Figure S6** shows that the Fermi level passes through the energy band, thereby indicating the metallic properties of this material. In addition, compared with the Fermi level of the undoped $CuIr_2Te_4$ ($E_F$ = 9.112 eV), the value of $E_F$ increased for $Cu_{0.9}Ag_{0.1}Ir_2Te_4$, thereby leading to an increase in $T_c$. Similar behavior has also been observed in a spinel system [60]. Therefore, we suggest that the initial increase in $T_c$ and disappearance of the CDW were due to the increase in the DOS at the Fermi surface, and the re-emergence of the CDW with higher Ag doping might have been the result of disorder scattering by Ag impurities and the decrease in the DOS at Fermi surface because some portions of the Fermi surface were detached by CDW gapping to decrease $T_c$ [56]. However, the mechanism responsible for the re-appearance of the CDW-like transition requires further investigation using low-temperature powder neutron diffraction, X-ray scattering, and high-resolution transmission electron microscopy methods at low temperature. The inset in **Fig. 6** shows the changes in the superconducting $T_c$ as the *c*/*a* ratio decreased. Clearly, the change in $T_c$ exhibited a dome-like shape as the unit cell increased (i.e., the *c*/*a* value), where $T_c$ reached a maximum value of 2.93 K when *c*/*a* = 1.3731, and the continuous increase in the unit cell as the Ag doping content increased caused the decrease in $T_c$.

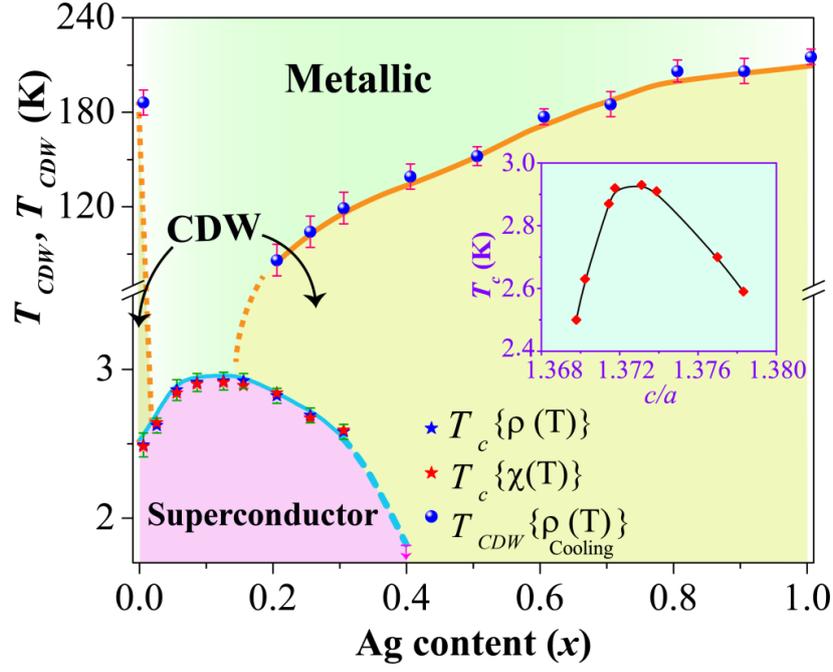

**Fig. 6** Electronic phase diagram for the system $Cu_{1-x}Ag_xIr_2Te_4$. $T_c$ values were determined from resistivity $\rho(T)$ and dc magnetic susceptibility ($\chi(T)$) measurements. $T_{CDW}$ values were determined from $d\rho/dT$.

**Table 3** Characteristic parameters for $Cu_{1-x}Ag_xIr_2Te_4$ compared with other telluride chalcogenide series.

| Material \ Parameter | $Cu_{0.95}Ag_{0.05}Ir_2Te_4$ | $Cu_{0.92}Ag_{0.08}Ir_2Te_4$ | $Cu_{0.88}Ag_{0.12}Ir_2Te_4$ | $Cu_{0.85}Ag_{0.15}Ir_2Te_4$ | $Cu_{0.8}Ag_{0.2}Ir_2Te_4$ | $CuIr_2Te_4$ [25] | $CuIr_{1.95}Ru_{0.05}Te_4$ [28] | $CuIr_2Te_{3.9}I_{0.1}$ [30] | $CuIr_2Te_{3.9}Se_{0.1}$ [31] | $Cu_{0.5}Zn_{0.5}Ir_2Te_4$ [27] | $CuIr_{1.95}Ti_{0.05}Te_4$ [51] |
|---|---|---|---|---|---|---|---|---|---|---|---|
| $T_c$ (K) | 2.87 | 2.92 | 2.93 | 2.91 | 2.8 | 2.5 | 2.79 | 2.95 | 2.83 | 2.82 | 2.84 |
| $\gamma$ (mJ/mol K$^2$) | | 12.1 | 13.9 | 14.8 | | 12.05 | 12.26 | 12.97 | 10.84 | 13.37 | 14.13 |
| $\beta$ (mJ/mol K$^4$) | | 1.98 | 2.12 | 2.08 | | 1.97 | 1.87 | 3.03 | 3.51 | 1.96 | 2.72 |
| $\Theta_D$ (K) | | 189 | 186 | 187 | | 190 | 193 | 165 | 157 | 191 | 171 |
| $\Delta C_{el}/\gamma T_c$ | | 1.40 | 1.44 | 1.42 | | 1.5 | 1.51 | 1.46 | 1.53 | 1.45 | 1.34 |
| $\lambda_{ep}$ | | 0.62 | 0.64 | 0.63 | | 0.63 | 0.65 | 0.70 | 0.65 | 0.66 | 0.64 |
| $N(E_F)$ (states/eV f.u.) | | 3.27 | 3.61 | 3.68 | | 3.1 | 3.15 | 3.24 | 3.11 | 3.41 | 3.67 |

| $H_{c1}(0)$ (Oe) | 181 | 154 | 135 | | 280 | 980 | 240 | 660 | 620 | 950 |
| $H_{c2}(0)$(Oe)($\rho_{N50\%}$) | | | 2100 | 1400 | 1200 | 2470 | 1880 | 1440 | 1980 | 2120 |
| $H^P$ (T) | | | 5.2 | 5.2 | 4.7 | 5.2 | 5.5 | 5.3 | 5.3 | 5.3 |
| $\xi_{GL}$ (nm) | | | 40 | 48 | 53 | 36 | 42 | 47 | 41 | 39 |

## 4. Conclusion

In this study, we successfully synthesized a series of polycrystalline $Cu_{1-x}Ag_xIr_2Te_4$ alloys where $0 \leq x \leq 1$. Powder XRD analysis clearly indicated that Ag substituted in the Cu site in $CuIr_2Te_4$ but without a structural phase transition. Temperature-dependent resistivity and magnetization analyses showed that the optimal $T_c$ value was obtained at $x = 0.12$ with a maximum of 2.93 K. However, the CDW order vanished rapidly when $x = 0.02$ and then suddenly re-appeared at $x = 0.2$, and $T_{CDW}$ increased as $x$ increased. In general, SC formed a dome-shaped area in the phase diagram linked by two CDW regions. These results provide insights into the mechanism responsible for the interaction between SC and the CDW.

**Acknowledgments**

This study was supported by the National Natural Science Foundation of China (Grant No. 11922415), Guangdong Basic and Applied Basic Research Foundation (2019A1515011718), Key Research & Development Program of Guangdong Province, China (2019B110209003), and the Pearl River Scholarship Program of Guangdong Province Universities and Colleges (20191001). Collection of the heat capacity data was supported by Guangdong Provincial Key Laboratory (Grant No. 2019B121203002).

**References**

[1] B. Keimer, S.A. Kivelson, M.R. Norman, S. Uchida, J. Zaanen, From quantum matter to high-temperature superconductivity in copper oxides, *Nature* **518** (7538) (2015) 179-186.
[2] G. Campi, A. Bianconi, N. Poccia, G. Bianconi, L. Barba, G. Arrighetti, D. Innocenti, J. Karpinski, N.D. Zhigadlo, S.M. Kazakov, M. Burghammer, M.v. Zimmermann, M. Sprung, A.

Ricci, Inhomogeneity of charge-density-wave order and quenched disorder in a high-$T_c$ superconductor, *Nature* **525**(7569) (2015) 359-362.

[3] P. Walmsley, C. Putzke, L. Malone, I. Guillamón, D. Vignolles, C. Proust, S. Badoux, A. I. Coldea, M. D. Watson, S. Kasahara, Y. Mizukami, T. Shibauchi, Y. Matsuda, A. Carrington, Quasiparticle mass enhancement close to the quantum critical point in BaFe$_2$(As$_{1-x}$P$_x$)$_2$, Physical Review Letters **110** (2013) 257002.

[4] T. Sarkar, D.S. Wei, J. Zhang, N.R. Poniatowski, P.R. Mandal, A. Kapitulnik, R.L. Greene, Ferromagnetic order beyond the superconducting dome in a cuprate superconductor, *Science* **368**(6490) (2020) 532-534.

[5] L. Taillefer, Scattering and Pairing in Cuprate Superconductors, *Annual Review of Condensed Matter Physics* **1**(1) (2010) 51-70.

[6] S. Badoux, W. Tabis, F. Laliberté, G. Grissonnanche, B. Vignolle, D. Vignolles, J. Béard, D. A. Bonn, W. N. Hardy, R. Liang, N. Doiron-Leyraud, L. Taillefer, and C. Proust, Change of carrier density at the pseudogap critical point of a cuprate superconductor, *Nature* **531** (2016) 210.

[7] K. Shimizu, T. Kimura, S. Furomoto, K. Takeda, K. Kontani, Y. Onuki, K. Amaya, Superconductivity in the non-magnetic state of iron under pressure, *Nature* **412** (6844) (2001) 316-318.

[8] S.C. Shen, B.B. Chen, H.X. Xue, G. Cao, C.J. Li, X.X. Wang, Y.P. Hong, G.P. Guo, R.F. Dou, C.M. Xiong, L. He, J.C. Nie, Gate dependence of upper critical field in superconducting (110) LaAlO$_3$/SrTiO$_3$ interface, *Scientific Reports* **6**(1) (2016) 28379.

[9] M. Thiemann, M.H. Beutel, M. Dressel, N.R. Lee-Hone, D.M. Broun, E. Fillis-Tsirakis, H. Boschker, J. Mannhart, M. Scheffler, Single-Gap Superconductivity and Dome of Superfluid Density in Nb-Doped SrTiO$_3$, *Physical Review Letters* **120**(23) (2018) 237002.

[10] Y. Tomioka, N. Shirakawa, K. Shibuya, I.H. Inoue, Enhanced superconductivity close to a non-magnetic quantum critical point in electron-doped strontium titanate, *nature Communications* **10**(1) (2019) 738.

[11] R. Peierls Quantum, *Theory of Solids* (1955). Oxford University Press, London.

[12] Y. Kvashnin, D. VanGennep, M. Mito, S. A. Medvedev, R. Thiyagarajan, O. Karis, A. N. Vasiliev, O. Eriksson, M. Abdel-Hafiez, Coexistence of Superconductivity and Charge Density

Waves in Tantalum Disulfide: Experiment and Theory, *Physical Review Letter* **125** (2020) 186401

[13] A. Majumdar, D. VanGennep, J. Brisbois, D.Chareev, A.V. Sadakov, A.S. Usoltsev, M. Mito, A.V. Silhanek, T. Sarkar, A. Hassan, O. Karis, R. Ahuja, M. Abdel-Hafiez, Interplay of charge density wave and multiband superconductivity in layered quasi-two-dimensional materials: The case of 2H−$NbS_2$ and 2H−$NbSe_2$, *Physical Review Materials* **4** (2020) 084005.

[14] YI. Joe, X.M. Chen, P. Ghaemi, K.D. Finkelstein, G.A. de la Peña, Y. Gan, J.C.T. Lee, S. Yuan, J. Geck, G.J. MacDougall, T.C. Chiang, S.L. Cooper, E. Fradkin, P. Abbamonte, Emergence of charge density wave domain walls above the superconducting dome in 1T-$TiSe_2$, *Nature Physics* **10**(6) (2014) 421-425.

[15] S. Kitou, A. Nakano, S. Kobayashi, K. Sugawara, N. Katayama, N. Maejima, A. Machida, T. Watanuki, K. Ichimura, S. Tanda, T. Nakamura, H. Sawa, Effect of Cu intercalation and pressure on excitonic interaction in 1T-$TiSe_2$, *Physical Review B* **99**(10) (2019) 104109.

[16] K. Cho, M. Kończykowski, S. Teknowijoyo, M.A. Tanatar, J. Guss, P.B. Gartin, J.M. Wilde, A. Kreyssig, R.J. McQueeney, A.I. Goldman, V. Mishra, P.J. Hirschfeld, R. Prozorov, Using controlled disorder to probe the interplay between charge order and superconductivity in $NbSe_2$, *Nature Communications* **9**(1) (2018) 2796.

[17] A.M. Gabovich, A.I. Voitenko, M. Ausloos, Charge- and spin-density waves in existing superconductors: competition between Cooper pairing and Peierls or excitonic instabilities, *Physics Reports* **367**(6) (2002) 583-709.

[18] H. Luo, T. Klimczuk, L. Müchler, L. Schoop, D. Hirai, M.K. Fuccillo, C. Felser, R.J. Cava, Superconductivity in the $Cu(Ir_{1-x}Pt_x)_2Se_4$ spinel, *Physical Review B* **87**(21) (2013) 214510.

[19] M. Monteverde, J. Lorenzana, P. Monceau, M. Núñez-Regueiro, Quantum critical point and superconducting dome in the pressure phase diagram of *o*-$TaS_3$, *Physical Review B* **88**(18) (2013) 180504.

[20] X.-L. Yu, J. Wu, Superconducting dome driven by intervalley phonon scattering in monolayer $MoS_2$, *New Journal of Physics* **22**(1) (2020) 013015.

[21] X.-C. Pan, X. Wang, F. Song, B. Wang, The study on quantum material $WTe_2$, *Advances in Physics: X* **3**(1) (2018) 1468279.

[22] X.-C. Pan, X. Chen, H. Liu, Y. Feng, Z. Wei, Y. Zhou, Z. Chi, L. Pi, F. Yen, F. Song, X. Wan, Z. Yang, B. Wang, G. Wang, Y. Zhang, Pressure-driven dome-shaped superconductivity and electronic structural evolution in tungsten ditelluride, *Nature Communications* **6**(1) (2015) 7805.

[23] L. Li, X. Deng, Z. Wang, Y. Liu, M. Abeykoon, E. Dooryhee, A. Tomic, Y. Huang, J.B. Warren, E.S. Bozin, S.J.L. Billinge, Y. Sun, Y. Zhu, G. Kotliar, C. Petrovic, Superconducting order from disorder in 2H-TaSe$_{2-x}$S$_x$, *npj Quantum Materials* **2**(1) (2017) 11.

[24] S. Koley, N. Mohanta, A. Taraphder, Charge density wave and superconductivity in transition metal dichalcogenides, *The European Physical Journal B* **93** (2020) 77.

[25] D. Yan, Y. Zeng, G. Wang, Y. Liu, J. Yin, T.-R. Chang, H. Lin, M. Wang, J. Ma, S. Jia, D.-X. Yao, H. Luo, CuIr$_2$Te$_4$: A Quasi-Two-Dimensional Ternary Telluride Chalcogenide Superconductor, 2019, p. arXiv:1908.05438.

[26] S. Nagata, N. Kijima, S. Ikeda, N. Matsumoto, R. Endoh, S. Chikazawa, I. Shimono, H. Nishihara, Resistance anomaly in CuIr$_2$Te$_4$, *Journal of Physics and Chemistry of Solids* **60**(2) (1999) 163-165.

[27] D. Yan, Y. Zeng, Y. Lin, L. Zeng, J. Yin, M. Boubeche, M. Wang, Y. Wang, D.-X. Yao, H. Luo, Robust Superconductivity in (Zn$_x$Cu$_{1-x}$)$_{0.5}$IrTe$_2$, *The Journal of Physical Chemistry C* **125**(10) (2021) 5732-5738.

[28] D. Yan, L. Zeng, Y. Lin, J. Yin, Y. He, X. Zhang, M. Huang, B. Shen, M. Wang, Y. Wang, D. Yao, H. Luo, Superconductivity in Ru-doped CuIr$_2$Te$_4$ telluride chalcogenide, *Physical Review B* **100**(17) (2019) 174504.

[29] D. Yan, Y. Zeng, Y. Lin, L. Zeng, J. Yin, Y. He, M. Boubeche, M. Wang, Y. Wang, D.-X. Yao, H. Luo, Charge density wave and superconductivity in the family of telluride chalcogenides Zn$_{1-x}$Cu$_x$Ir$_{2-y}$N(N = Al, Ti, Rh)$_y$Te$_4$, 2020, p. arXiv:2003.11463.

[30] M. Boubeche, J. Yu, L. Chushan, W. Huichao, L. Zeng, Y. He, X. Wang, W. Su, M. Wang, D.-X. Yao, Z. Wang, H. Luo, Superconductivity and charge density wave in iodine-doped CuIr$_2$Te$_4$, *Chinese Physics Letters* **38**(3) (2021) 037401.

[31] M. Boubeche, N. Wang, J. Sun, P. Yang, L. Zeng, Q. Li, Y. He, S. Luo, J. Cheng, y. peng, H. Luo, Anomalous Charge Density Wave State Evolution and Dome-like Superconductivity

in CuIr$_2$Te$_{4-x}$Se$_x$ Chalcogenides, *Superconductor Science and Technology* **34**(11) (2021)115003.

[32] E Nazarova, N Balchev, K Nenkov, K Buchkov, D Kovacheva, A Zahariev, G Fuchs, Transport and pinning properties of Ag-doped FeSe$_{0.94}$, *Superconductor Science & Technology* **28** (2015) 025013.

[33] Y.C. Guo, H.K. Liu and SX Dou, Silver-doped (Bi,Pb)$_2$Sr$_2$Ca$_2$Cu$_3$O$_{10}$/Ag high-temperature superconducting composites, *Physica C* **215** (1993) 291-296.

[34] P. E. Blöchl, Projector augmented-wave method. *Phys. Rev. B* **50** (1994) 17953-17979.

[35] G. Kresse, D. Joubert, From ultrasoft pseudopotentials to the projector augmented-wave method. *Phys. Rev. B* **59** (1999) 1758-1775.

[36] H.M. Rietveld, The Rietveld Method? A Historical Perspective, *Australian Journal of Physics* **41**(2) (1988) 113-116.

[37] X-L. Wang, S.X. Dou, M.S.A. Hossain, Z.X. Cheng, X.Z. Liao, S.R. Ghorbani, Q.W. Yao, J.H. Kim, T. Silver, Enhancement of the in-field $J_c$ of MgB$_2$ via SiCl$_4$ doping, *Physical Review B* **81** (2010) 224514.

[38] KSB De Silva, X. Xu, X.L. Wang, D. Wexler, D. Attard, F. Xiang, S.X. Dou A significant improvement in the superconducting properties of MgB$_2$ by co-doping with graphene and nano-SiC, S*cripta Materialia* **67**(10) (2012) 802-5.

[39] H.T. Wang, L.J. Li, DS. Ye, X.H. Cheng, Z.A. Xu, Effect of Te doping on superconductivity and charge-density wave in dichalcogenides 2H-NbSe$_{2-x}$Te$_x$ ($x$ = 0, 0.1, 0.2) *Chinese Physics* **16** (2007) 2471.

[40] E. Morosan, H.W. Zandbergen, B.S. Dennis, J.W.G. Bos, Y. Onose, T. Klimczuk, A. P. Ramirez, N.P. Ong, R.J Cava, Superconductivity in Cu$_x$TiSe$_2$, *Nature Physics* **2**(8) (2006) 544-50.

[41] T. Baumgartner, Effects of fast neutron irradiation on critical currents and intrinsic properties of state-of-the-art Nb$_3$Sn wires, *Vienna University of Technology*, 2013.

[42] T. Baumgartner, M. Eisterer, H.W. Weber, R. Flükiger, C. Scheuerlein, L. Bottura, Effects of neutron irradiation on pinning force scaling in state-of-the-art Nb$_3$Sn wires, *Superconductor Science & Technology* **27**(1) (2013) 015005.

[43] E. Helfand, N.R. Werthamer, Temperature and purity dependence of the superconducting Critical Field $H_{c2}$, *Physical Review* **147**(1) (1966) 288-294.

[44] M. Tinkham, Introduction to Superconductivity. United States, *Dover Publications* (2004).

[45] V.L. Ginzburg, On the theory of superconductivity, *Il Nuovo Cimento* (1955-1965) **2**(6) (1955) 1234-1250.

[46] T. McConville, B. Serin, Ginzburg-Landau parameters of type-II superconductors, *Physical Review* **140** (4A) (1965) A1169-A1177.

[47] T. Gruner, D. Jang, Z. Huesges, R. Cardoso-Gil, G. H. Fecher, M. M. Koza, O. Stockert, A. P. Mackenzie, M. Brando, and C. Geibel, Charge density wave quantum critical point with strong enhancement of superconductivity, *Nature Physics* **13** (10) (2017) 967-972.

[48] H. Cao, B.C. Chakoumakos, X. Chen, J. Yan, M.A McGuire, H. Yang, R. Custelcean, H. Zhou, D.J. Singh, D. Mandrus, Origin of the phase transition in $IrTe_2$: Structural modulation and local bonding instability, *Physical Review B* **88** (2013) 115122.

[49] K. Kudo, M. Kobayashi, S. Pyon, M. Nohara, Suppression of structural phase transition in $IrTe_2$ by isovalent Rh doping, *Journal of the Physical Society of Japan* **82** (2013) 085001.

[50] K. Jin, K. Liu, What drives superconductivity in Pt-doped $IrTe_2$?, *Science Bulletin* **60** (2015) 822-822.

[51] L. Zeng, D. Yan, Y. He, M. Boubeche, Y. Huang, X. Wang, H. Luo, Effect of Ti substitution on the superconductivity of $CuIr_2Te_4$ telluride chalcogenide, *Journal of Alloys and Compounds* **885** (2021)160981.

[52] G.A. Scholz, Charge-density-wave behaviour in intercalated single crystal $Nb_3Te_4$, *Solid State Ionics* **100**(1) (1997) 135-141.

[53] N.V. Selezneva, E.M. Sherokalova, V.G. Pleshchev, V.A. Kazantsev, N.V. Baranov, Suppression and inducement of the charge-density-wave state in $Cr_xTiSe_2$, *Journal of Physics: Condensed Matter* **28**(31) (2016) 315401.

[54] NV Baranov, V.I. Maksimov, J. Mesot, V.G. Pleschov, A. Podlesnyak, V. Pomjakushin, N.V. Selezneva, Possible re-appearance of the charge density wave transition in $M_xTiSe_2$ compounds intercalated with 3$d$ metals, *Journal of Physics: Condensed Matter* **19**(1) (2006) 016005.

[55] Y. Liu, R. Ang, W.J. Lu, W.H. Song, LJ Li, YP Sun, Superconductivity induced by Se-doping in layered charge-density-wave system 1T-TaS$_{2-x}$Se$_x$, *Applied Physics Letters* **102**(19) (2013) 192602.

[56] J.M. Harper, T.E. Geballem, F.J. Disalvo, Thermal properties of layered transition-metal dichalcogenides at charge-density-wave transitions, *Physical Review B* **15** (1977) 2943-2951.

[57] R.L. Barnett, A. Polkovnikov, E. Demler, W.G. Yin, W. Ku, Coexistence of gapless excitations and commensurate charge-density wave in the 2H transition metal dichalcogenides, *Physical Review Letters* **96** (2006) 026406.

[58] P.W. Anderson, Knight Shift in Superconductors, *Physical Review Letters* 3 (1959) 325.

[59] X. Zhu, B. Lv, F. Wei, Y. Xue, B. Lorenz, L. Deng, Y. Sun, C-W. Chu, Disorder-induced bulk superconductivity in ZrTe$_3$ single crystals via growth control, *Physical Review B* 87(2) (2013) 024508.

[60] H. X. Luo, T. Klimczuk, L. Muchler, L. Schoop, D. Hirai, M. K. Fuccillo, C. Felser, and R. J. Cava, Superconductivity in the Cu(Ir$_{1-x}$Pt$_x$)$_2$Se$_4$ spinel, *Physical Review B* **87** (2013) 214510.

# Supplemental information

# Enhanced superconductivity with possible re-appearance of charge density wave states in polycrystalline $Cu_{1-x}Ag_xIr_2Te_4$ alloys

Mebrouka Boubeche[a,#], Lingyong Zeng[a,#], Xunwu Hu[b], Shu Guo[c,d], Yiyi He[a], Peifeng Yu[a], Yanhao Huang[a], Chao Zhang[a], Shaojuan Luo[e], Dao-Xin Yao[b], Huixia Luo[a*]

[a] School of Materials Science and Engineering, State Key Laboratory of Optoelectronic Materials and Technologies, Key Lab of Polymer Composite & Functional Materials, Guangzhou Key Laboratory of Flexible Electronic Materials and Wearable Devices, Sun Yat-Sen University, No. 135, Xingang Xi Road, Guangzhou, 510275, P. R. China

[b] School of Physics, State Key Laboratory of Optoelectronic Materials and Technologies, Sun Yat-Sen University, No. 135, Xingang Xi Road, Guangzhou, 510275, P. R. China

[c] Shenzhen Institute for Quantum Science and Engineering, Southern University of Science and Technology, Shenzhen 518055, China.

[d] International Quantum Academy, Shenzhen 518048, China.

[e] School of Chemical Engineering and Light Industry, Guangdong University of Technology, Guangzhou, 510006, P. R. China

[#] These authors contributed equally to this work.
[*] Author to whom any correspondence should be addressed.
E-mail: luohx7@mail.sysu.edu.cn

**Table S1** Rietveld refinement structural parameters of $Cu_{1-x}Ag_xIr_2Te_4$ with the $P$-$3m$1 space group (No. 164).

| **$Cu_{0.98}Ag_{0.02}Ir_2Te_4$** | | | | $R_{wp}$ = 3.88%, $R_p$ = 3.02%, $R_{exp}$ = 2.13%, $\chi^2$ = 3.51 | | |
|---|---|---|---|---|---|---|
| **Label** | **x** | **y** | **z** | **site** | **Occupancy** | **Multiplicity** |
| Cu | 0 | 0 | 0.5 | 2$b$ | 0.49 | 1 |
| Ag | 0 | 0 | 0.5 | 2$b$ | 0.01 | 1 |
| Ir | 0 | 0 | 0 | 1$a$ | 1 | 1 |
| Te | 0.33333 | 0.66667 | 0.73967(9) | 2$b$ | 1 | 2 |
| **$Cu_{0.95}Ag_{0.05}Ir_2Te_4$** | | | | $R_{wp}$ = 3.68%, $R_p$ = 3.01%, $R_{exp}$ = 2.02%, $\chi^2$ = 3.32 | | |
| **Label** | **x** | **y** | **z** | **site** | **Occupancy** | **Multiplicity** |
| Cu | 0 | 0 | 0.5 | 2$b$ | 0.475 | 1 |
| Ag | 0 | 0 | 0.5 | 2$b$ | 0.025 | 1 |
| Ir | 0 | 0 | 0 | 1$a$ | 1 | 1 |
| Te | 0.33333 | 0.66667 | 0.74402(3) | 2$b$ | 1 | 2 |
| **$Cu_{0.92}Ag_{0.08}Ir_2Te_4$** | | | | $R_{wp}$ = 3.18%, $R_p$ = 3.05%, $R_{exp}$ = 2.05%, $\chi^2$ = 3.12 | | |
| **Label** | **x** | **y** | **z** | **site** | **Occupancy** | **Multiplicity** |
| Cu | 0 | 0 | 0.5 | 2$b$ | 0.46 | 1 |
| Ag | 0 | 0 | 0.5 | 2$b$ | 0.04 | 1 |
| Ir | 0 | 0 | 0 | 1$a$ | 1 | 1 |
| Te | 0.33333 | 0.66667 | 0.74422(2) | 2$b$ | 1 | 2 |
| **$Cu_{0.85}Ag_{0.15}Ir_2Te_4$** | | | | $R_{wp}$ = 3.34%, $R_p$ = 2.91%, $R_{exp}$ = 2.06%, $\chi^2$ =3.5 | | |
| **Label** | **x** | **y** | **z** | **site** | **Occupancy** | **Multiplicity** |
| Cu | 0 | 0 | 0.5 | 2$b$ | 0.375 | 1 |
| Ag | 0 | 0 | 0.5 | 2$b$ | 0.125 | 1 |
| Ir | 0 | 0 | 0 | 1$a$ | 1 | 1 |
| Te | 0.33330 | 0.66667 | 0.74541(8) | 2$b$ | 1 | 2 |
| **$Cu_{0.8}Ag_{0.2}Ir_2Te_4$** | | | | $R_{wp}$ = 3.4 %, $R_p$ = 2.92%, $R_{exp}$ = 2.1%, $\chi^2$ = 3.52 | | |
| **Label** | **x** | **y** | **z** | **site** | **Occupancy** | **Multiplicity** |
| Cu | 0 | 0 | 0.5 | 2$b$ | 0.1 | 1 |
| Ag | 0 | 0 | 0.5 | 2$b$ | 0.4 | 1 |
| Ir | 0 | 0 | 0 | 1$a$ | 1 | 1 |
| Te | 0.33330 | 0.66667 | 0.74572(6) | 2$b$ | 1 | 2 |
| **$Cu_{0.75}Ag_{0.25}Ir_2Te_4$** | | | | $R_{wp}$ = 3.38 %, $R_p$ = 2.35%, $R_{exp}$ = 2.09%, $\chi^2$ =3.48 | | |
| **Label** | **x** | **y** | **z** | **site** | **Occupancy** | **Multiplicity** |
| Cu | 0 | 0 | 0.5 | 2$b$ | 0.375 | 1 |
| Ag | 0 | 0 | 0.5 | 2$b$ | 0.125 | 1 |
| Ir | 0 | 0 | 0 | 1$a$ | 1 | 1 |
| Te | 0.33330 | 0.66667 | 0.74589(4) | 2$b$ | 1 | 2 |
| **$Cu_{0.7}Ag_{0.3}Ir_2Te_4$** | | | | $R_{wp}$ = 3.54 %, $R_p$ = 2.45%, $R_{exp}$ = 2.1%, $\chi^2$ =3.54 | | |
| **Label** | **x** | **y** | **z** | **site** | **Occupancy** | **Multiplicity** |
| Cu | 0 | 0 | 0.5 | 2$b$ | 0.35 | 1 |
| Ag | 0 | 0 | 0.5 | 2$b$ | 0.15 | 1 |
| Ir | 0 | 0 | 0 | 1$a$ | 1 | 1 |

| Te | 0.33330 | 0.66667 | 0.74618(9) | 2b | 1 | 2 |

| **$Cu_{0.6}Ag_{0.4}Ir_2Te_4$** | | | $R_{wp}$ = 3.59 %, $R_p$ = 2.56%, $R_{exp}$ = 2.11%, $\chi^2$ = 3.54 | | | |
|---|---|---|---|---|---|---|
| **Label** | **x** | **y** | **z** | **site** | **Occupancy** | **Multiplicity** |
| Cu | 0 | 0 | 0.5 | 2b | 0.3 | 0.5 |
| Ag | 0 | 0 | 0.5 | 2b | 0.2 | 0.5 |
| Ir | 0 | 0 | 0 | 1a | 1 | 1 |
| Te | 0.33330 | 0.66667 | 0.74627(9) | 2b | 1 | 2 |

| **$Cu_{0.4}Ag_{0.6}Ir_2Te_4$** | | | $R_{wp}$= 3.53 %, $R_p$= 2.44%, $R_{exp}$ = 2.13%, $\chi^2$ = 3.43 | | | |
|---|---|---|---|---|---|---|
| **Label** | **x** | **y** | **z** | **site** | **Occupancy** | **Multiplicity** |
| Cu | 0 | 0 | 0.5 | 2b | 0.2 | 1 |
| Ag | 0 | 0 | 0.5 | 2b | 0.3 | 1 |
| Ir | 0 | 0 | 0 | 1a | 1 | 1 |
| Te | 0.33330 | 0.66667 | 0.74682(9) | 2b | 1 | 2 |

| **$Cu_{0.3}Ag_{0.7}Ir_2Te_4$** | | | $R_{wp}$ = 3.75 %, $R_p$ = 2.95%, $R_{exp}$ = 2.15%, $\chi^2$ = 3.28 | | | |
|---|---|---|---|---|---|---|
| **Label** | **x** | **y** | **z** | **site** | **Occupancy** | **Multiplicity** |
| Cu | 0 | 0 | 0.5 | 2b | 0.25 | 1 |
| Ag | 0 | 0 | 0.5 | 2b | 0.35 | 1 |
| Ir | 0 | 0 | 0 | 1a | 1 | 1 |
| Te | 0.33330 | 0.66667 | 0.74827(3) | 2b | 1 | 2 |

| **$Cu_{0.2}Ag_{0.8}Ir_2Te_4$** | | | $R_{wp}$ = 3.21 %, $R_p$ = 3.13%, $R_{exp}$ = 2.13%, $\chi^2$ =3.22 | | | |
|---|---|---|---|---|---|---|
| **Label** | **x** | **y** | **z** | **site** | **Occupancy** | **Multiplicity** |
| Cu | 0 | 0 | 0.5 | 2b | 0.1 | 1 |
| Ag | 0 | 0 | 0.5 | 2b | 0.4 | 1 |
| Ir | 0 | 0 | 0 | 1a | 1 | 1 |
| Te | 0.33330 | 0.66667 | 0.74923(4) | 2b | 1 | 2 |

| **$Cu_{0.1}Ag_{0.9}Ir_2Te_4$** | | | $R_{wp}$ = 3.39 %, $R_p$ = 3.15%, $R_{exp}$ = 2.04%, $\chi^2$ = 3.35 | | | |
|---|---|---|---|---|---|---|
| **Label** | **x** | **y** | **z** | **site** | **Occupancy** | **Multiplicity** |
| Cu | 0 | 0 | 0.5 | 2b | 0.05 | 1 |
| Ag | 0 | 0 | 0.5 | 2b | 0.45 | 1 |
| Ir | 0 | 0 | 0 | 1a | 1 | 1 |
| Te | 0.33330 | 0.66667 | 0.74923(6) | 2b | 1 | 2 |

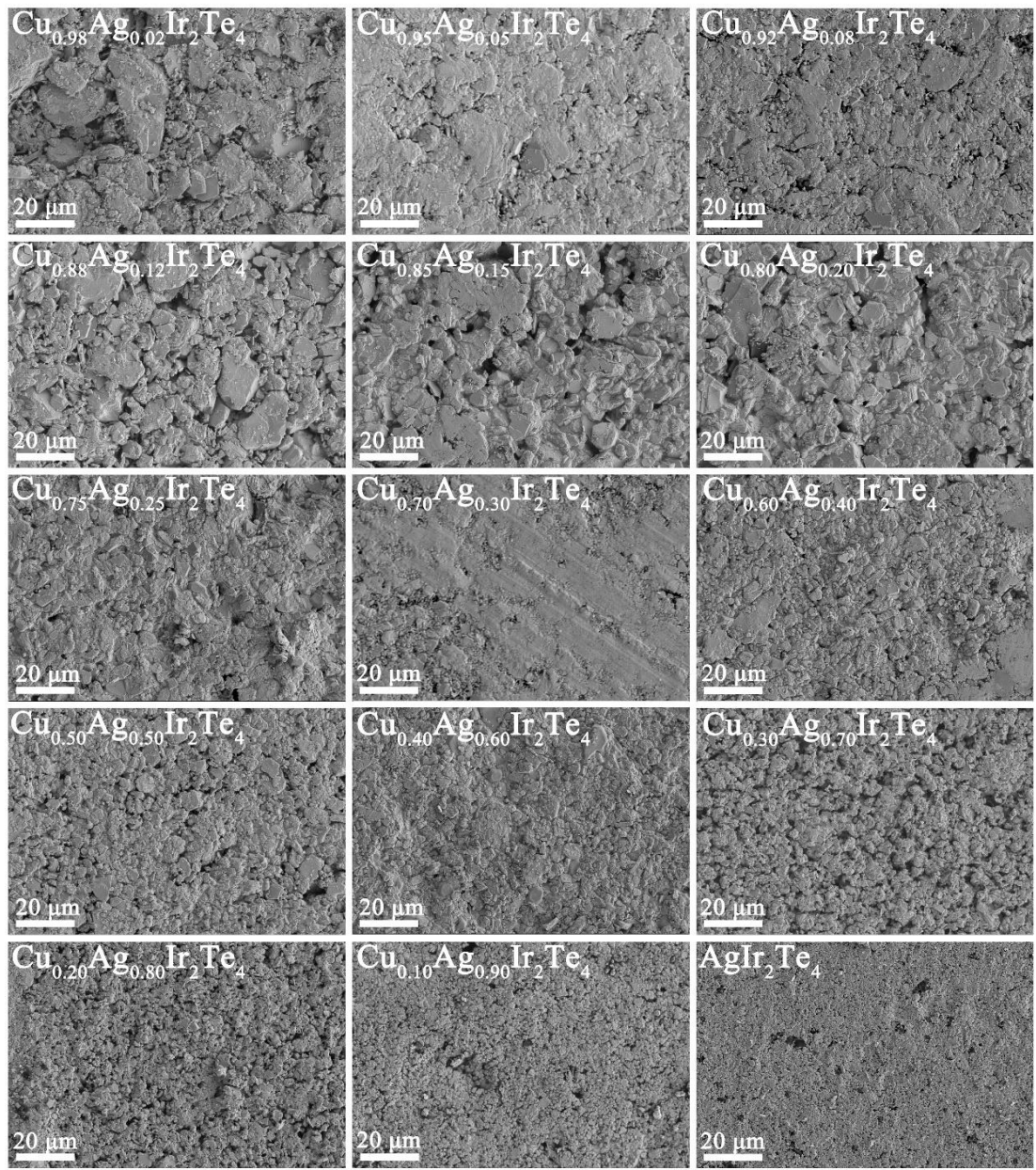

**Fig. S1** SEM images of $Cu_{1-x}Ag_xIr_2Te_4$ samples.

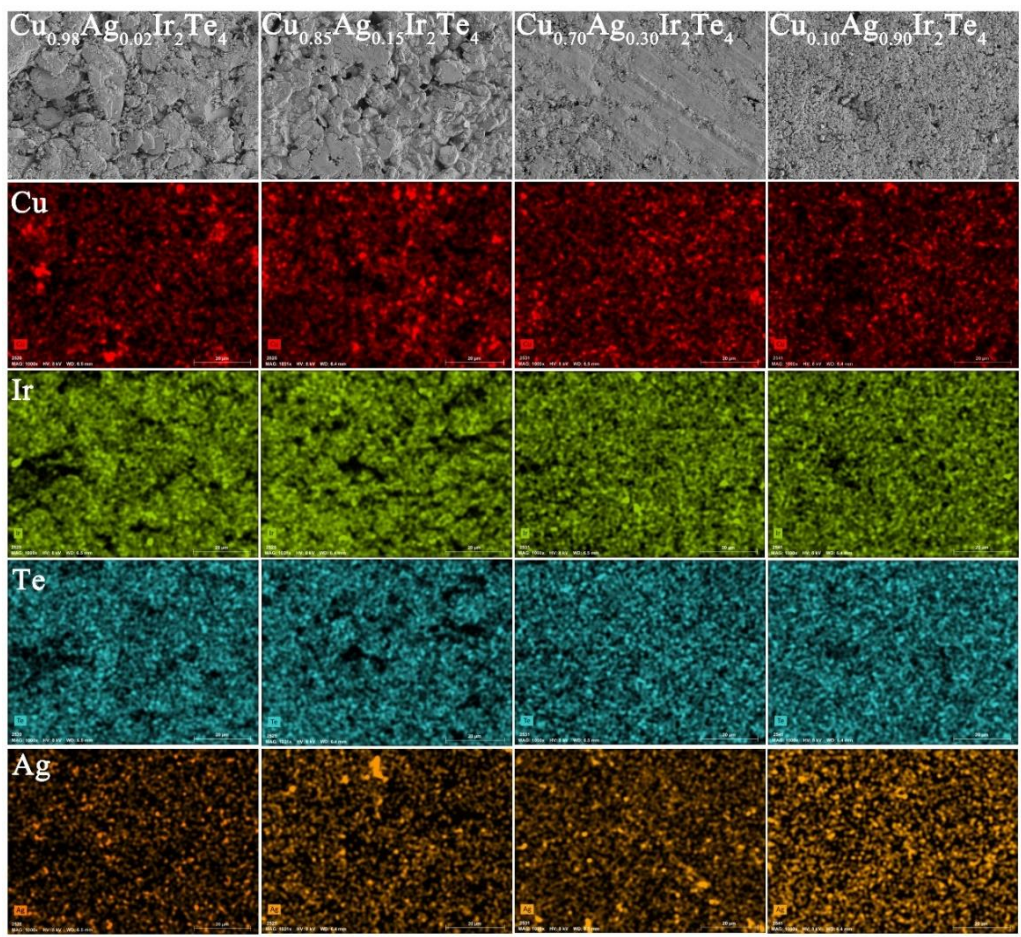

**Fig. S2** EDXS mappings of Cu$_{1-x}$Ag$_x$Ir$_2$Te$_4$ polycrystalline samples.

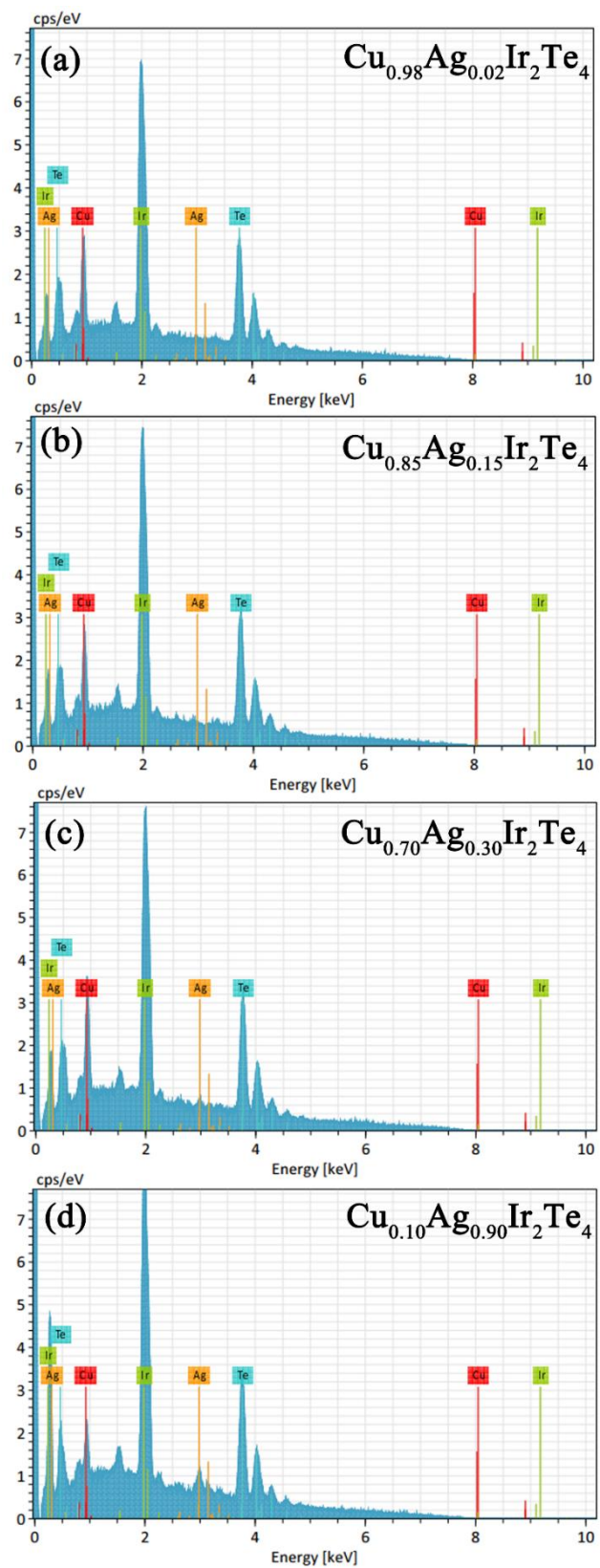

**Fig. S3** EDXS spectrum of $Cu_{1-x}Ag_xIr_2Te_4$.

**Table S2:** The element ratios of $Cu_{1-x}Ag_xIr_2Te_4$ from EDXS results.

| Element ratio Sample | Cu | Ir | Te | Ag |
|---|---|---|---|---|
| **$CuIr_2Te_4$** [1] | 0.97 | 1.96 | 3.93 | 0 |
| **$Cu_{0.98}Ag_{0.02}Ir_2Te_4$** | 0.94 | 1.96 | 4.04 | 0.017 |
| **$Cu_{0.85}Ag_{0.15}Ir_2Te_4$** | 0.89 | 1.95 | 4.07 | 0.127 |
| **$Cu_{0.70}Ag_{0.30}Ir_2Te_4$** | 0.76 | 1.97 | 3.96 | 0.269 |
| **$Cu_{0.10}Ag_{0.90}Ir_2Te_4$** | 0.13 | 1.92 | 4.06 | 0.867 |

[1] D. Yan, Y. Zeng, G. Wang, Y. Liu, J. Yin, T.-R. Chang, H. Lin, M. Wang, J. Ma, S. Jia, D.-X. Yao, H. Luo, $CuIr_2Te_4$: A Quasi-Two-Dimensional Ternary Telluride Chalcogenide Superconductor, 2019, p. arXiv:1908.05438.

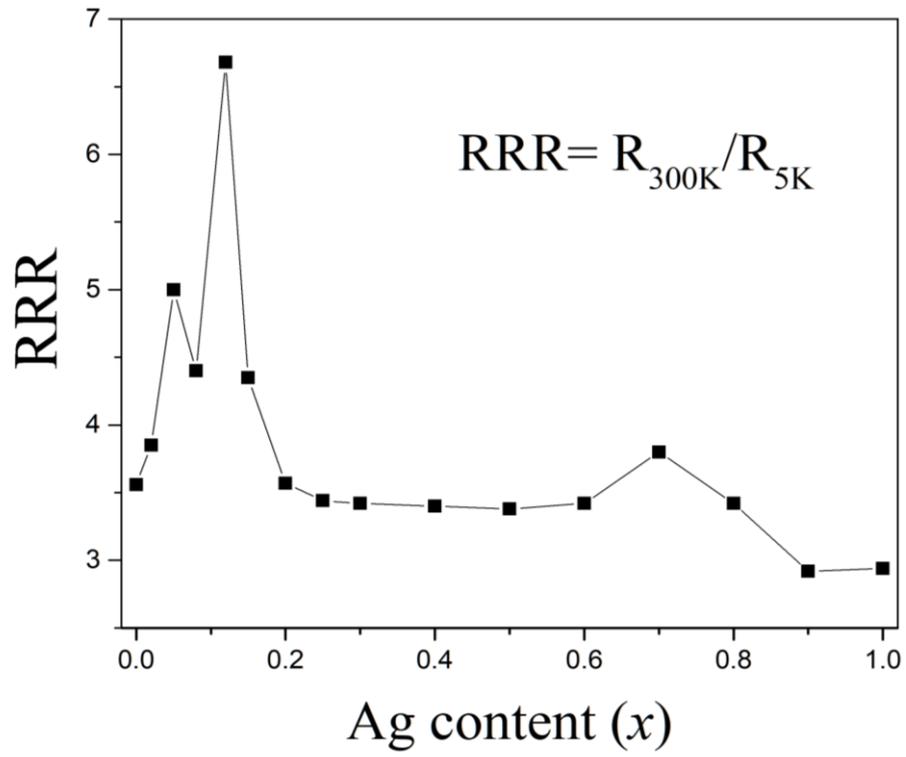

**Fig. S4.** Doping dependent residual resistivity ratio (RRR) for $Cu_{1-x}Ag_xIr_2Te_4$.

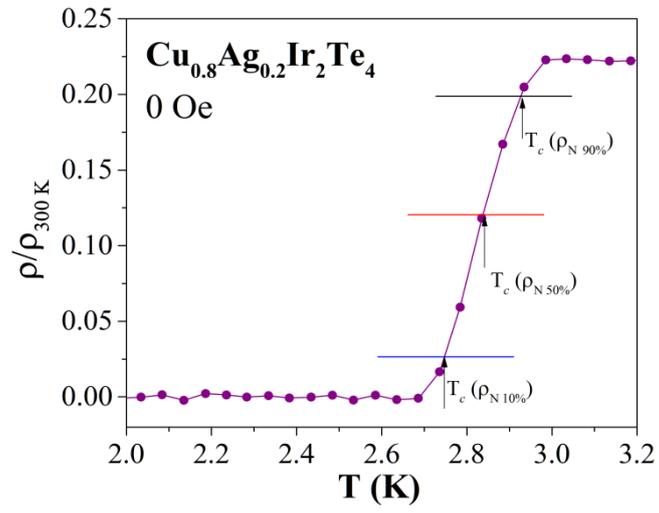

**Fig. S5.** Definition of the criteria $\rho_N$ 90%, $\rho_N$ 50%, $\rho_N$ 10%

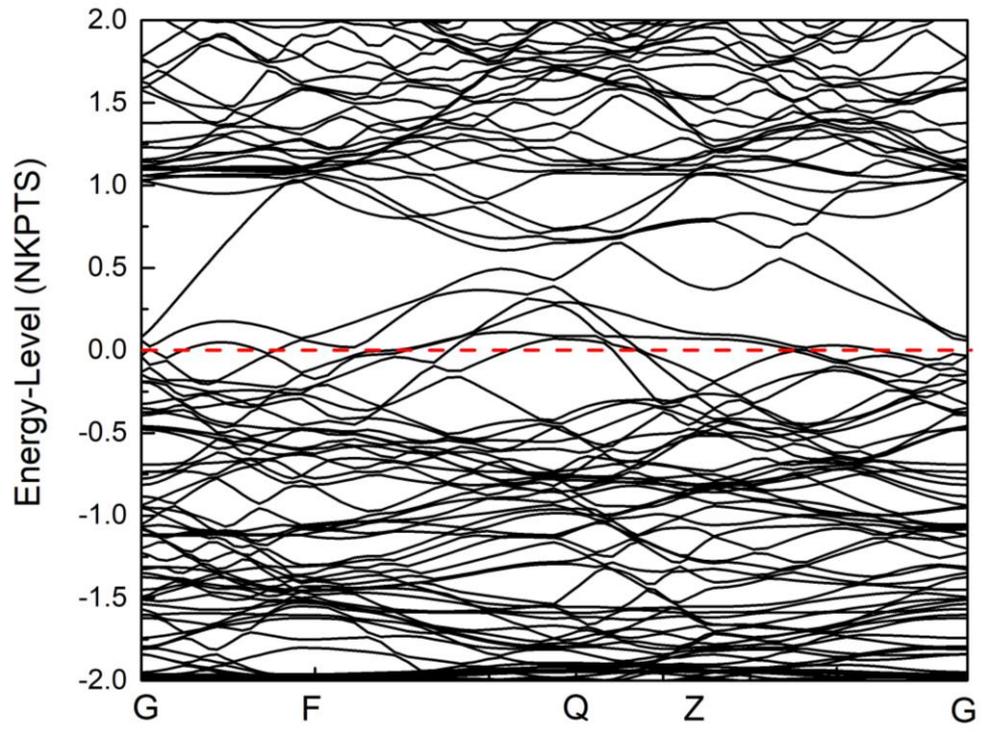

**Fig. S6.** Calculated band structure of $Cu_{0.9}Ag_{0.1}Ir_2Te_4$